\pdfoutput=1
\documentclass[sigplan,10pt,nonacm=true]{acmart}

\makeatletter                   
\def\mdseries@tt{m}             
\makeatother        

\makeatletter
\renewcommand\@formatdoi[1]{\ignorespaces}
\makeatother

\usepackage[export]{adjustbox}
\usepackage{listings}                                                                
\usepackage[newfloat=true,frozencache=true]{minted} 
\usepackage{color}                                                                   
\usepackage{xcolor}                                                                  
\usepackage{changepage} 
                                                                                     
\usepackage{url}                                                                     
\hypersetup{hidelinks=false,colorlinks=true,linkcolor=blue,urlcolor=blue,citecolor=blue,anchorcolor=blue} 

\usepackage{microtype}                                                               
                                                                                     
\usepackage[iso,american,cleanlook]{isodate}    

\usepackage{rviewport}                                                               

\usepackage[inline]{enumitem}

\newcommand\Invisible[1]{                                                            
  \marginpar{\color{white}{\fontsize{.5}{.5}\selectfont #1 }}                        
}

\newcommand{\Exclude}[1]{}

\newcommand\Boldly[1]{\vspace{0.5 \baselineskip} \noindent \textbf{$\blacktriangleright$} \textbf{#1} \noindent}

\definecolor{Gray95}{gray}{0.95}
                                                                                     
\newcommand{\AtFoot}[1]{\let\thefootnote\relax\footnotetext{{#1}}}

\acmConference[]{}{}{}{}
\acmISBN{}
\acmDOI{}
\startPage{1} 

\setcopyright{none}

\usepackage{booktabs}   
\usepackage{subcaption} 

\newcommand{\orcidicon}[1]{\href{https://orcid.org/#1}{\includegraphics[scale=0.06]{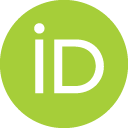}}}

\begin{document}

\title[]{BRAVO -- Biased Locking for Reader-Writer Locks} 


\author{Dave Dice \orcidicon{0000-0001-9164-7747}} 
\orcid{0000-0001-9164-7747}             
\affiliation{
  \institution{Oracle Labs}             
}
\email{first.last@oracle.com}           

\author{Alex Kogan \orcidicon{0000-0002-4419-4340}}
\orcid{0000-0002-4419-4340}            
\affiliation{
  \institution{Oracle Labs}             
}
\email{first.last@oracle.com}         

\affiliation{
}


\begin{abstract}

Designers of modern reader-writer locks confront a difficult trade-off related to reader scalability.
Lock implementations that have a compact memory representation for active readers will typically 
suffer under high intensity read-dominated workloads when the ``reader indicator'' state is updated
frequently by a diverse set of threads, causing cache invalidation and coherence traffic.  
Other designs use distributed reader indicators, one per NUMA node, per core or even per thread.
This improves reader-reader scalability, but also increases the size of each lock 
instance and creates overhead for writers.  


We propose a simple transformation, \underline{BRAVO}, that augments any existing reader-writer lock, adding 
just two integer fields to the lock instance.  Readers make their presence known to writers by hashing 
their thread's identity with the lock address, forming an index into a \emph{visible readers table} and
installing the lock address into the table.
All locks and threads in an address space can share the same readers table.  
Crucially, readers of the same lock tend to write to different locations in the table, reducing coherence traffic.
Therefore, BRAVO can augment a simple compact lock to provide scalable concurrent reading,
but with only modest and constant increase in footprint.

We implemented BRAVO in user-space, as well as integrated it with the Linux kernel reader-writer semaphore (\texttt{rwsem}).
Our evaluation with numerous benchmarks and real applications, both in user and kernel-space,
demonstrate that BRAVO improves performance and scalability of underlying locks 
in read-heavy workloads while introducing virtually no overhead, including in workloads in
which writes are frequent.

\Invisible{Readers attempt to install the lock address into that element in the table, 
making their existence known to potential writers.  
Updates by readers tend to be diffused over the table, resulting in a NUMA-friendly design.  
Specifically, BRAVO allows a simple compact lock to be augmented so as to provide scalable concurrent reading
but with only a modest increase in footprint.} 

\end{abstract}




\maketitle

\thispagestyle{fancy}

\section{Introduction}

A reader-writer lock, also known as a shared-exclusive lock, is a
synchronization primitive for controlling access by multiple threads (or processes)
to a shared resource (critical section). 
It allows shared access for read-only use of the resource, while write operations
access the resource exclusively. 
Such locks are ubiquitous in modern systems, and can be found, for example,
in database software, file systems, key-value stores and operating systems. 

Reader-writer locks have to keep track of the presence of 
active readers before a writer can be granted the lock.
In the common case, such presence is recorded in a shared counter, incremented and
decremented with every acquisition and release of the lock in the read mode.
This is the way reader-writer locks are implemented in the Linux kernel, POSIX pthread library
and several other designs~\cite{rtsj10-brandenburg, spaa12-shirako, ppopp91-Mellor-Crummey}.
The use of a shared counter lends itself to a relatively simple implementation and 
has a compact memory representation for a lock.
However, it suffers under high intensity read-dominated workloads when the ``reader indicator'' state is updated
frequently by a diverse set of threads, causing cache invalidation and coherence traffic~\cite{CB08, usenixatc14-liu, spaa13-dice,europar13-dice}. 

Alternative designs for reader-writer locks use distributed reader indicators,  
for instance, one per NUMA node as in cohort locks~\cite{ppopp13-calciu},
or even one lock per core as in the Linux kernel brlock~\cite{brlock2} 
and other related ideas~\cite{ipps92-hsieh,vyukov,distributedrw,usenixatc14-liu}.
This improves reader-reader scalability, but also considerably increases the size of each lock 
instance.
Furthermore, the lock performance is hampered when writes are frequent, as
multiple indicators have to be accessed and/or modified.
Finally, such locks have to be instantiated dynamically, since
the number of sockets or cores can be unknown until the runtime.
As a result, designers of modern reader-writer locks confront a difficult trade-off related to 
the scalability of maintaining the indication of the readers' presence.

In this paper, we propose a simple transformation, called BRAVO, that augments any existing 
reader-writer lock, adding just two integer fields to the lock instance.
When applied on top of a counter-based reader-writer lock, BRAVO
allows us to achieve, and often beat, the performance levels of locks that 
use distributed reader indicators while maintaining a compact footprint of the underlying lock.
With BRAVO, readers make their presence known to writers by hashing their thread's identity 
with the lock address, forming an index into a \emph{visible readers table}. 
Readers attempt to install the lock address into the element (slot) in the table identified by that index.
If successful, readers can proceed with their critical section \emph{without} modifying
the shared state of the underlying lock.
Otherwise, readers resort to the acquisition path of the underlying lock.
Note that the visible readers table is shared by all locks and threads in an address space.  
Crucially, readers of the same lock tend to write to different locations in the table, 
reducing coherence traffic and thus resulting in a NUMA-friendly design.
At the same time, a writer always uses the acquisition path of the underlying lock, 
but also scans the readers table and waits for all readers that acquired that lock through it.
A simple mechanism is put in place to limit the overhead of scanning the table
for workloads in which writes are frequent.

We implemented BRAVO and evaluated it on top of several locks, 
such as the POSIX pthread\_rwlock lock and the PF-Q reader-writer lock by 
Brandenburg and Anderson~\cite{rtsj10-brandenburg}.
For our evaluation, we used numerous microbenchmarks as 
well as \texttt{rocksdb}~\cite{rocksdb}, a popular open-source key-value store.
Furthermore, we integrated BRAVO with \texttt{rwsem}, a read-write 
semaphore in the Linux kernel.
We evaluated the modified kernel through kernel microbenchmarks
as well as several user-space applications (from the Metis suite~\cite{MMK10}) that create contention on 
read-write semaphores in the kernel.
All our experiments in user-space and in the kernel demonstrate that BRAVO 
is highly efficient in improving performance and scalability of underlying locks 
in read-heavy workloads while introducing virtually no overhead, even in workloads in
which writes are frequent.

The rest of the paper is organized as follows.
The related work is surveyed in Section~\ref{sec:related}.
We present the BRAVO algorithm in Section~\ref{sec:bravo} and discuss how
we apply BRAVO in the Linux kernel in Section~\ref{sec:rwsem}.
The performance evaluation in user-space and the Linux kernel is provided in Sections~\ref{sec:user-space-evaluation}
and~\ref{sec:kernel-evaluation}, respectively.
We conclude the paper and elaborate on multiple directions for future work in Section~\ref{sec:conclusions}.




\Invisible{
*  single-writer multiple-reader locks
*  Readers must typically write to shared data to announce their action and make themselves 
   visible to writers, but ideally, in order to reduce coherence traffic, avoid writing to 
   locations frequently read or written by other readers.
*  Announce; publicize; publish; make visible; 
*  Tension; trade-off
*  Primum non-nocere ; bound harm; hippocratic oath
*  Collision probability in VRT is equivalent to "Birthday Paradox".
*  Dual paths for readers : fast and slow
   alternate fast-path encoding/representation/encoding
*  Cost model : improvement = BenefitFromFastReaders - RevocationCost
   Unfortunately the BenefitFromFastReaders is a function of the number of readers.
*  Improvement ratio = performance(BRAVO-A) / performance(A) 
*  revocation occurs on transition/edge from fast read to write.  
*  dispersed; diffused; distributed; disseminate; split; fractured; sharded; 
   decomposed; dilute; spread; 
*  deconstruct; decouple; 
*  faithful; realistic; accurate; fidelity; veracity; authentic; adherent; 
*  Insufficiently algorithmic
*  A short idea expressed in a long paper
*  Accelerator layer; stack; compose; transform; construct; fabricate; wrap
*  wrapper; jacket; envelope
*  subsume; subducts
*  Downsides : adaptivity impacts performance predictability; 
*  trade-off : table size vs collision rate and revoke scan overhead
*  shift some cost/burden from readers to writers
*  performance predictability; variability; variance; consistent; 
*  adhere to principle of least surprise
*  divert; revert; fall-back; fail-over
*  desiderata; performance goals; target; ideal; aspire; 
*  devolve; degenerate; converge; trend toward; tend towards
*  slipstream; 
*  confers
*  affords
*  encumbrance
*  defensible decision/design
*  supplant
*  consequent; ensuant; ensue; pursuant; arising; by virtue of; 
*  constrain; limit; guard; clamp; guard; cap; protect; restrict; 
*  throttle downside; stop-loss; stop-limit
*  performance diode
*  reader-writer VS read-write VS RW 
*  compact "inverted" encoding : collapse 2 fields into one
   change RBias to Inhibit
   InhibitUntil = 0 indicates bias enabled
   InhibitUntil = T indicates bias disabled until time T
   Possibly use T=MaxValue for disabled
*  release; surrender; relinquish; unlock; abjure; 
*  Thought experiment : Gedankenexperiment
   Imagine reenable reader-bias on every read, or after every write
*  Additional aspect to cost model
   Penalties associated with bias enabled
   @  revocation cost
   @  collisions in fast-read attempt -- futile CAS
   @  sharing or false-sharing from near-collisions in table
*  Conservative and pessimistic -- Loss Aversion
   Trade-off biased toward loss aversion instead 
   We forsake/forgo/surrender potentially better performance in order to limit 
   the possibility of worsened performance.
*  target environment for BRAVO :
   read-dominated
   multiple frequently arriving concurrent readers
*  assiduously avoid
*  relax 
*  Topology oblivious; topology insensitive; topology independent
*  BA has neutral preference
*  Databases will commonly use arrays of reader-writer locks, with 
   concurrency protection sharded via hash function.
*  revoke on write-after-fast-read 
*  Minimize
*  Inter-node miss vs intra-node coherence miss
*  Augment and extent
*  race; inopportune interleaving; intervened; window
*  Slot; Array element; Table element; cell; index; 
*  Writers always pass through the underlying lock.
*  Incremental cost
*  See BRAVO.txt for additional commentary and discussion
*  in-effect; prevailing preference; transparent; 
*  reader in fast-path = fast reader
*  We note that if reader-writer lock algorithm \emph{A} has certain preference properties 
   (reader preference, writer preference, neutral preference, etc) then \emph{BRAVO-A} will 
   exhibit the same properties.  BRAVO is ``transparent'' and does not alter the admission policy 
   of the underlying lock. 
*  slow-down bound in general vs slow-down bound specifically for writers
*  Fast-path readers do not write into the lock instance, only the table.
*  When necessary writers must scan 
   the array to check for conflicting readers, so our approach is most useful for read-dominated workloads. 
*  read-read concurrency
*  Numerous; myriad; plurality; 
*  without surrendering ...; trade-off
*  Profligate memory consumption -- footprint
*  Use hash function to scatter accesses into hash table
*  Stochastic disperal
*  Probabalistically confict-free or conflict-reducing
*  Reader arrival : write to dispersed RI but then read from central WI
*  NUMA vs NUCA
*  NUMA-friendly because write sharing is particularly expensive on NUMA
*  Common read-write locks with central RI generates write-sharing coherence traffic.
   BRAVO avoids that central RI
*  Oracle ID accession number ORA190125
*  Confounding factor : high pthread\_rwlock\_t diversity -- large number of different lock 
   instances in use over a short period -- can result in increased cache pressure arising 
   from accesses into the BRAVO visible readers table.  
   Increased cache footprint and pressure arising from accesses to array/table
*  Claim : regarding collisions and near collisions, concurrency is 
   equivalent to lock diversity -- no difference.  

TODO:

*  modality : favor read-dominated; read mostly
*  As shown, writers in revocation phase spin busy-wait, but it's relatively trivial
   to make them wait politely via spin-then-park instead. 
*  Equivalent ski-rental problem
*  VRT sizing : should be a function of the number of logical CPUs.  
   Want at least one cache sector per CPU.  
*  TODO: report read/write fraction and fast-read fraction = NFast / (NFast+NSlow) 
   FastReads; SlowReads; WriteNormal; WriteRevoke;
   SlowRead breakdown : SlowReadDiabled; SlowReadCollision
*  Briefly, we have a simple wrapper that lets you augment any existing reader-writer lock, and 
   yields a composite lock that will have good read-read scaling.
*  BRAVO-2D : To improve temporal locality, divide the visible readers table in sectors, each with, 
   say, 256 contiguous slots.
   We use the caller's CPUID to identify a sector, and then a hash function on the lock address to identify
   a slot within that sector.  This allows a secondary optimization where revoking threads need to
   check only one location -- the index identified by the hash on the lock address -- 
   in each sector during revocation. 
*  AKA : BRAVO-2D; BRAVO-Sectored; BRAVO-RowColumn; 
}

\Invisible{
*  avoid futile setting where bias likely to be revoked in short order
*  improved polices and more faithful cost model for bias setting
*  Implement BRAVO over mutex, but may deny expected R-R admission.  
*  explore interplay between preference of underlying lock and BRAVO.  Writer preference locks 
   appear to provide benefit when used in conjunction with BRAVO as writer preferences reduces the
   read-to-write transition rate, and thus and revocation costs, instead grouping writers together so only the
   lead writer is required to revoke bias.
*  We believe our approach may be applicable to user-mode RCU
}


\Invisible{

Clarify nuanced and subtle points about inter-lock interference 

Statement-1

*  Interference does affect performance, but in a ``reasonable'' way.  
   While some interference exists, BRAVO still provides benefit and naturally it 
   is most efficient when one or only a few locks are accessed since in the case of multiple 
   locks, each of those locks is less contended.  

*  Convexity observed in inter-lock interference sensitivity plot
   at N=small BRAVO does very well relative to BRAVO-BA-Prime as there is no inter-lock interference.
   At N=Large BRAVO also does relatively well because contention is dispersed and diffused over many locks.
   At N=middling we some some evidence of inter-lock interference.  

*  If we hold the number of threads constant and increase the number of locks, then we increase
   the number of distinct slots accessed in the visible readers table -- occupancy increases. 
   In turn, the odds and rate of collisions or near-collisions then increases accordingly.  
   But the odds of a specific given lock encountering contention also decrease as we have
   increased lock diversity.  Thus, for a given lock L, fewer threads are forced into
   the slow path for L at any given time.  As fewer threads use the slow path for L, we reduce
   coherence traffic on L's central reader indicator. 
   Given the reduced traffic into the slow path, the cost of using the slow path for L is not 
   excessive so the penalty of using the slow path is mitigated.   
   As we increase N, the consequent increased aggregated collision rate is offset by faster 
   slow path operations.  

*  The slow path is not really "slow" if only one thread uses the path at a given time.  

*  More collisions but lower intensity in the slow path, and thus faster slow path execution,
   mitigating those collisions.  

Statement-2 :

*  Normal execution mode vs mode with very small number of distinct locks

*  Thought experiment : 
   model BRAVO running concurrent benchmark where each thread picks locks randomly 
   from a pool of N.  
   model where threads run in lockstep -- in-phase stepping
   Concurrently, each thread picks a random lock, acquires it, and then releases it.
   This is equivalent to threads throwing ball into bins where balls
   are lock acquisitions and bins are indices in the visible readers table.  
   We assume the hash function provides equidistribution and is effectively
   equivalent to randomization of indices. 
   Can use balls-into-bins probability model to analyze collisions in the table.
   Collision rate per access is Balls / (2*Bins).
   The number of locks is NOT relevant to the collision rate.

*  Claim and conjecture
   In general the collision rate in the readers table is purely a function of just the 
   tablesize and the number of concurrent threads and NOT the number of distinct locks.

*  Increasing the number of locks does not normally increase inter-lock interference effects.

*  Increasing the number of locks may influence the rate of near collisions and 
   related coherence traffic.

*  Increasing the number of locks influences cache/TLB pressure and footprint, 
   potentially resulting in capacity misses.

*  BUT : consider the operating region where we have a small number of locks N in use.
   The set of locks and threads may map to only a subset of the table via the hash. 
   With small N, the effective set of reachable bins is smaller.  
   As we increase the number of locks, a larger fraction of the table is accessed.
   Occupancy increases.  
   This increases the inter-lock collision rate -- inter-lock interference.
   In this region, as we increase N we increase the inter-lock collision rate. 
   But if we hold the number of threads constant and increase N, the 
   odds of a conflict on a given lock L will drop because of increased diversity. 
   In turn the slow path for L is used less frequently, making it less expensive
   offseting the cost of the collision.  

*  BRAVO yields a very good performance-space trade-off, which is the key point.
   (INCLUDE in CONCLUSION) 

*  BRAVO offers a viable alternative that resides between centralized BA with small
   footprint and poor reader scalability and the large locks with good readers scalability.  
   (INCLUDE in CONCLUSION) 

*  BRAVO yields a favorable performance trade-off between space and scalability,  
   offering a viable alternative that resides on the design spectrum between classic 
   centralized locks, such as BA, having small footprint and poor reader scalability, 
   and the large locks with high reader scalability.

Statement-3

USE THE FOLLOWING TEXT in the inter-lock interference section.  

Lets assume as usual that we’re holding the number of threads and the size of the BRAVO 
visible readers table fixed, and that we’re varying the number of locks N.  
Threads randomly pick locks for a shared pool of N and acquire and release “read” permission.    

When N is relatively small, the range of the hash function — mapping (thread X lock) space 
to table indices -- hits just a subset of the table.  The effective size of the table -- 
reachable indices -- is smaller than the actual size.   As we increase N this effect abates, 
the range increases, we diffuse over more of the table, and the true collision rate decreases.  
(Increasing N further reaches an asymptote, and the true collision rate devolves into a 
simple balls-into-bins collision model.  That is, at some N the true collision rate is solely 
a function of the number of thread and the size of the table, and N is no longer relevant).   
However, as we increase N, the range of indices where a given thread can "hit" -- the diversity -- 
also increases, and we incur more false sharing.    
Furthermore we can encounter increased "temporal sharing".  
Say thread T previously accessed index I, and then accesses I again.  As we increase N, the odds 
that there have been no intervening accesses to I by other threads will decrease, and T’s 
access will miss and incur additional coherence traffic.  (Temporal sharing is sharing that 
is not a concurrent conflict and is not false sharing).    In a sense, as we increase N, 
we cause additional writing to slots that were more recently written by other threads.  
Locality decreases and cache misses increase.  Again, there’s an asymptote here, after which 
performance is insensitive to further increases in N, as the hash fully diffuses accesses for a 
given lock over the table.  

In addition, as we increase N, the odds decrease that when thread T accesses index I, that 
T was the last thread to access that I.   That is, the odds decrease that there were no 
intervening accesses to I by other threads since T last accessed I.   

Remarks:

*  Sensitivity analysis -- Supplementary table and experiments  which supports our 
   claim that inter-lock interference is not a signficant performance issue.  
*  See : InterferenceRW.cc and BRAVOModel*.cc and BallsIntoBins.cc
*  Consider: run sensitivity survey varying TableSize and N, 
   and identify worst case BRAVO/LARGE ratio and report that.
   LARGE refers to the family locks with large footprint that provide scalable 
   concurrent reading. 
   Includes : DV-BA; BRAVO-BA-Prime and PrimePrime; Cohort locks; etc
*  Assume equidistribution of hash function that makes (Thread,Lock) to index in table. 
   Equivalent to randomization so we can use balls-into-bins probability model. 
*  Collision rate insensitive to number of locks and inter-lock interference.  
*  true collision : results in fall-back into slow path and futile coherence traffic
   near collision : results in false sharing and coherence traffic
*  Keywords : 
   @  Diversity; distribute; spread; diffuse; 
   @  occupancy; load; utilization; saturation; tenancy; residency; 
   @  intensity; arrival rate
*  Explication : divide into 2 execution modes -- operating regions
   full table vs partial table  
*  N is very small implies effective number of reachable bins is small.
   only use subset of table.  
*  Our hash function may be better than random with respect to dispersal
    
}

\section{Related Work}
\label{sec:related} 

\Boldly{Reader-Indicator Design} Readers that are \emph{active} -- currently executing
in a reader critical section -- must be visible to potential writers.  Writers must
be able to detect active readers in order to resolve read-vs-write conflicts, and wait for
active readers to depart.  The mechanism through which readers make themselves visible
is the \emph{reader indicator}.  Myriad designs have been described in the literature.  
At one end of the spectrum we find a centralized reader indicator implemented as an integer field 
within each reader-writer lock instance that reflects the number of active readers.  
Readers use atomic instructions (or a central lock) to safely increment and decrement this field.
Classic examples of such locks can be found in the early work of Mellor-Crummey and Scott~\cite{ppopp91-Mellor-Crummey}
and more recent work by Shirako et al.~\cite{spaa12-shirako}.  
Another reader-writer lock algorithm having a compact centralized reader indicator is Brandenburg and Anderson's 
Phase-Fair Ticket lock, designated \texttt{PF-T} in \cite{rtsj10-brandenburg}, where
the reader indicator is encoded in two central fields.  
Their Phase-Fair Queue-based lock, \texttt{PF-Q}, uses a centralized counter for active readers 
and an MCS-like central queue, with local spinning, for readers that must wait.   
We refer to this latter algorithm as ``BA'' throughout the remainder of this paper.  
Such approaches are compact, having a small per-lock footprint, and simple, but, because of coherence traffic,
do not scale in the presence of concurrent readers that are arriving and departing
frequently~\cite{spaa13-dice,europar13-dice, CB08, usenixatc14-liu}.  

To address this concern, many designs turn toward distributed reader indicators.  
Each cohort reader-writer lock~\cite{ppopp13-calciu}, for instance, uses a per-NUMA node 
reader indicator.  While distributed reader indicators improve scalability,
they also significantly increase the footprint of a lock instance, with each reader indicator residing
on its own private cache line or sector to reduce false sharing. 
In addition, the size of the lock is variable with the number of nodes, 
and not known at compile-time, precluding simple static preallocation of locks.  
Writers are also burdened with the overhead of checking multiple 
reader indicators.
Kashyap et al.~\cite{usenixatc17-kashyap} attempt to address some of those issues by maintaining a dynamic list of 
per-socket structures and expand the lock instance on-demand.
However, this only helps if a lock is accessed by threads running on a subset of nodes.


At the extreme end of the spectrum we find lock designs with reader indicators assigned per-CPU 
or per-thread \cite{ipps92-hsieh,brlock2,vyukov,distributedrw,usenixatc14-liu}.  These designs promote read-read 
scaling, but have a large variable-sized footprint.  They also favor readers in that writers must 
traverse and examine all the reader-indicators to resolve read-vs-write conflicts, possibly 
imposing a performance burden on writers.  We note there are a number of varieties of such distributed locks : 
a set of reader-indicators coupled with a central mutual exclusion lock for writer permission, as found
in cohort locks~\cite{ppopp13-calciu}; 
sets of mutexes where readers must acquire one mutex and writers must acquire all mutexes,
as found in Linux kernel brlocks~\cite{brlock2}; 
or sets of reader-writer locks where readers must acquire read permission on one lock, and writers must acquire 
write permission on all locks.  
To reduce the impact on writers, which must visit all reader indicators,
some designs use a tree of distributed counters
where the root element contains a sum of the indicators within the subtrees~\cite{spaa09-lev}. 

Dice et al.~\cite{spaa10-tlrw} devised \emph{read-write byte-locks} for use in the \emph{TLRW} 
software transactional memory infrastructure.  Briefly, read-write byte-locks are reader-writer 
locks augmented with an array of bytes, serving as reader indicators,  where indices in the array
are assigned to favored threads that are frequent readers.  These threads can simply set and clear 
these reader indicators with normal store operations.  The motivation for read-write byte-locks was 
to avoid atomic read-modify-write instructions, which were particularly expensive on the system under test.  
The design, as described, is not NUMA-friendly as the byte array occupies a single cache line.  

In addition to distributing or dispersing the counters, individual counters can 
themselves be further split into constituent \texttt{ingress} and \texttt{egress}
fields to further reduce write sharing.  Arriving readers increment the ingress field
and departing readers increment the egress field.  Cohort reader-writer locks use this approach~\cite{ppopp13-calciu}. 

BRAVO takes a different approach, opportunistically representing active readers in the shared global 
visible readers table.  The table (array) is fixed in size and shared over all threads and locks within
an address space.  Each BRAVO lock has, in addition to the underlying reader-writer lock, a boolean
flag that indicates if reader bias is currently enabled for that lock.  Publication of active 
readers in the array is strictly optional and best-effort.  A reader can always fall back to acquiring
read permission via the underlying reader-writer lock.  
BRAVO’s benefit comes from reduced coherence traffic arising from reader arrival.
Such coherence traffic is particularly costly on NUMA systems, consuming shared interconnect 
bandwidth and also exhibiting high latency.  
As such, BRAVO is naturally NUMA-friendly, but
unlike most other NUMA-aware reader-writer locks, it does not need to understand or otherwise 
query the system topology,
further simplifying the design and reducing dependencies\footnote{While
BRAVO is topology oblivious, it does require high-resolution
low-latency means of reading the system clock.  We further expect that reading the clock is
scalable, and that concurrent readers do not interfere with each other.  On systems with 
modern Intel CPUs and Linux kernels the \texttt{RDTSCP} instruction or 
\texttt{clock\_gettime(CLOCK\_MONOTONIC)} fast system call suffices.}.
We note that coherence traffic caused by waiting — such as global vs local waiting — is 
determined by the nature of the underlying lock.

\Boldly{Optimistic Invisible Readers}
Synchronization constructs such as \emph{seqlocks} \cite{sosp71-easton,seqlock1,seqlock2} allow 
concurrent readers, but forgo the need for readers to make themselves visible.  Critically,
readers do not write to synchronization data and thus do not induce coherence traffic.
Instead, writers update state -- typically a modification counter -- to indicate that updates 
have occurred.  Readers check that counter at the start and then again at the end of their
critical section, and if writers were active or the counter changed, the readers self-abort and retry.  
An additional challenge for seqlocks is that readers can observe inconsistent state, and
special care must be taken to constrain the effects and avoid errant behavior in readers.  
Often, non-trivial reader critical sections must be modified to safely tolerate optimistic execution.  
Various hybrid forms exist, such as the \texttt{StampedLock}\cite{StampedLock} facility 
in \texttt{java.\allowbreak{}util.\allowbreak{}concurrent} which
consists of a reader writer lock coupled with a seqlock, providing 3 modes : classic pessimistic write locking, 
classic pessimistic read locking, and optimistic reading. 

To avoid the problem where optimistic readers might see inconsistent state, transactional lock elision
\cite{asplos09-dice,micro01-rajwar,tle-glibc,eurosys16-felber, DKL16} based on hardware transactional memory can be used.  
Readers are invisible and do not write to shared data.  Such approaches can be helpful, 
but are still vulnerable to indefinite abort and progress failure.  In addition, the 
hardware transactional memory facilities required to support lock elision are not available 
on all systems, and are usually best-effort, without any guaranteed 
progress, requiring some type of fallback to pessimistic mechanisms. 

\Boldly{Biased Locking}
BRAVO draws inspiration from \emph{biased locking}\cite{BiasedLocking, oopsla06-russell,pact10-vasudevan,pppj11-pizlo,QRL}. 
Briefly, biased locking allows the same thread to repeatedly acquire and release a mutual exclusion lock 
without requiring atomic instructions, except on the initial acquisition.  
If another thread attempted to acquire the lock, then 
expensive \emph{revocation} is required to wrest bias from the original thread.   
The lock would then revert to normal non-biased mode for some period before again becoming
potentially eligible for bias.  (Conceptually, we can think of the lock as just being left
in the locked state until there is contention. Subsequent lock and unlock operations by the original
thread are ignored -- the unlock operation is deferred until contention arises).  
Biased locking was a response to the CPU-local latencies incurred by atomic instructions on 
early Intel and SPARC processors and to the fact that locks in Java were often dominated 
by a single thread.  Subsequently, processor designers have addressed the latency concern, 
rendering biased locking less profitable.  

Classic biased locking identifies a preferred thread, while BRAVO identifies a preferred
access mode.  That is, BRAVO biases toward a mode instead of thread identity.
BRAVO is suitable for read-dominated workloads, 
allowing a fast-path for readers when reader bias is enabled for a lock.  If a write request 
is issued against a reader-biased lock, reader bias is disabled and revocation (scanning 
of the visible readers table) is required, 
shifting some cost from readers to writers.  Classic biased locking provides benefit by 
reducing the number of atomic operations and improving latency.  It does not improve scalability.  
BRAVO reader-bias, however, can improve both latency and scalability by reducing coherence 
traffic on the reader indicators in the underlying reader-writer lock.  

\section{The BRAVO Algorithm} 
\label{sec:bravo} 

BRAVO transforms any existing reader-writer lock \emph{A} into \emph{BRAVO-A}, which provides
scalable reader acquisition.  We say \emph{A} is the 
\emph{underlying} lock in \emph{BRAVO-A}.  In typical circumstances \emph{A} might be a simple compact lock that 
suffers under high levels of reader concurrency.  \emph{BRAVO-A} will also be compact, but is 
NUMA-friendly as it reduce coherence traffic and offers scalability in the presence of frequently 
arriving concurrent readers.  

Listing~\ref{Listing:BRAVO-py} depicts a pseudo-code implementation of the BRAVO algorithm.
BRAVO extends \emph{A}'s structure with a new \texttt{RBias} boolean field (Line~2).
Arriving readers first check the \texttt{RBias} field, and, if found set, then hash the 
address of the lock with a value reflecting the calling thread's identity to form an index
into the visible readers table (Lines~12--13).  
(This readers table is shared by all locks and threads in an
address space.  In all our experiments we sized the table at 4096 entries.  Each table element,
or \emph{slot}, is either \texttt{null} or a pointer to a reader-writer lock instance).  
The reader then uses an atomic compare-and-swap (CAS) operator to attempt to change the element
at that index from \texttt{null} to the address of the lock, publishing its existence to potential 
writers (Line~14).  If the CAS is successful then 
the reader rechecks the \texttt{RBias} field to ensure it remains set (Line~18).  If so, the reader has
successfully gained read permission and can enter the critical section (Line~19).  Upon completing the critical section
the reader executes the complementary operation to release read permission, simply storing
\texttt{null} into that slot (Lines~29--31).  We refer to this as the \emph{fast-path}.
The fast-path attempt prefix (Lines~11-23) runs in constant time.  
Our hash function is based on the \emph{Mix32} operator found in \cite{oopsla14-steele}.  

If the recheck operation above happens to fail, as would be the case if a writer intervened and
cleared \texttt{RBias} and the reader lost the race, then the reader simply clears the slot (Line~21) and reverts
to the traditional \emph{slow-path} where it acquires read permission via the underlying lock (Line~24).  
Similarly, if the initial check of \texttt{RBias} found the flag clear (Line~12), or the CAS failed
because of collisions in the array (Line~14) -- the slot was found to be populated -- then control diverts 
to the traditional slow-path.  After a slow-path reader acquires read permission from the underlying 
lock, it enters and executes the critical section, and then at unlock time releases read permission 
via the underlying lock (Line~33).  

\begin{listing}[htp]         
\begin{adjustwidth}{2em}{0pt}
\begin{minted}{py}

class BRAVOLock<T> : 
  int RBias 
  Time InhibitUntil
  T Underlying 

## Shared global :
BRAVOLock * VisibleReaders [4096] 
int N = 9   # slow-down guard 

def Reader(BRAVOLock * L) :
  BRAVOLock * * slot = null
  if L.RBias :
    slot = VisibleReaders + Hash(L, Self) 
    if CAS(slot, null, L) == null :
      # CAS succeeded
      # store-load fence required on TSO 
      # typically subsumed by CAS
      if L.RBias :        # recheck
        goto EnterCS      # fast path
      *slot = null        # raced - RBias changed
    slot = null
  # Slow path
  assert slot == null 
  AcquireRead (L.Underlying)
  if L.RBias == 0 and Time() >= L.InhibitUntil : 
    L.RBias = 1
  EnterCS:
  ReaderCriticalSection()
  if slot != null :
    assert *slot == L
    *slot = null
  else :
    ReleaseRead (L.Underlying)
 
def Writer(BRAVOLock * L) :
  AcquireWrite (L.Underlying)
  if L.RBias : 
    # revoke bias
    # store-load fence required on TSO
    L.RBias = 0 
    auto start = Time() 
    for i in xrange(VisibleReaders) : 
      while VisibleReaders[i] == L : 
        Pause() 
    auto now = Time() 
    # primum non-nocere : 
    # limit and bound slow-down
    # arising from revocation overheads
    L.InhibitUntil = now + ((now - start) * N) 
  WriterCriticalSection() 
  ReleaseWrite (L.Underlying) 
\end{minted} 
\vspace{-0.5cm} 
\captionof{listing}{Simplified Python-like Implementation of BRAVO\label{Listing:BRAVO-py}}
\end{adjustwidth}
\end{listing} 

Arriving writers first acquire write permission on the underlying reader-writer lock (Line~36).
Having done so, they then check the \texttt{RBias} flag (Line~37).  If set, the writer must perform
\emph{revocation}, first clearing the \texttt{RBias} flag (Line~40) and then scanning all the elements of
the visible readers table checking for conflicting fast-path readers (Lines~42--44).  If any elements 
match the lock, the writer must wait for that fast-path reader to depart and clear the slot.  
If lock $L$ has 2 fast-path active readers, for instance, then $L$ will appear twice in the array.  
Scanning the array might appear to be onerous, but in practice the
sequential scan is assisted by the automatic hardware prefetchers present in modern CPUs.  We observe
a scan rate of about 1.1 nanoseconds per element on our system-under-test (described later).   
Having checked \texttt{RBias} and performed revocation if necessary,
the writer then enters the critical section (Line~50).  At unlock-time, the writer simply releases write
permission on the underlying reader-writer lock (Line~51).  Therefore the only difference for writers under BRAVO is the 
requirement to check and potentially revoke reader bias if \texttt{RBias} was found set.
\Invisible{Amortized scan rate} 

We note that writers only scan the visible reader table, and never write into it.
Yet, this scan may pollute the writer's cache.
One way to cope with it is to use non-temporal loads, however, exploring this idea is left for the future work.
Note that revocation is only required on transitions from reading to writing and
only when \texttt{RBias} was previously set.

In summary, active readers can make their existence public in one of two ways : either 
via the visible readers table (fast-path), or via the traditional underlying reader-writer
lock (slow-path).  Our mechanism allows both slow-path and fast-path readers simultaneously.
Absent hash collisions, concurrent fast-path readers will write to different locations in the visible
readers table.  Collisions are benign, and impact performance but not correctness.
Writers resolve read-vs-write conflicts against fast-path readers via the visible readers table 
and against slow-path readers via the underlying reader-writer lock.

\Invisible{BRAVO provides a dual existence representation for active readers, with their existence 
reflected in either the array or the underlying lock.} 

One important remaining question is how to set \texttt{RBias}.
In our early prototypes we set \texttt{RBias} in the reader slow-path based on a low-cost Bernoulli trial 
with probability $P = 1/100$ using a thread-local Marsgalia XOR-Shift~\cite{jss03-marsaglia} 
pseudo-random number generator.  
While this simplistic policy for enabling bias worked well in practice, we were concerned
about situations where we might have enabled bias too eagerly, and incur frequent revocation to the point where
\emph{BRAVO-A} might be slower than \emph{A}.  Specifically, the worst-case scenario would be where slow readers
repeatedly set \texttt{RBias}, only to have it revoked immediately by a writer.  

The key additional cost in BRAVO is the revocation step, which
executes under the underlying write lock and thus serializes operations associated with the lock%
\footnote{Additional costs associated with BRAVO include futile atomic operations from collisions, 
and sharing or false-sharing arising from near-collisions in the table.  Our simplified cost model 
ignores these secondary factors. We note that the odds of collision are equivalent to those
given by the ``Birthday Paradox'' \cite{BirthdayParadox} and that the general problem of deciding
to set bias is equivalent to the classic ``ski-rental'' problem \cite{SkiRental}.}.  
As such, we measure the latency of revocation and multiply that period by $N$, a configurable parameter,
and then inhibit
the subsequent setting of bias in the reader slow-path for that period, bounding the worst-case
expected  slow-down from BRAVO for writers to $1/(N+1)$ (cf.~Lines~41-49). Our specific performance goal 
is \emph{primum non nocere} -- first, do no harm, with \emph{BRAVO-A} never 
underperforming \emph{A} by any significant margin on any workload\footnote{
Our approach conservatively forgoes the potential of better performance afforded
by the aggressive use of reader bias in order to limit the possibility
of worsened performance\cite{LossAversion}.}.  
This tactic is simple and effective, but excessively
conservative, taking into account only the worst-case performance penalty imposed by BRAVO, 
and not accounting for any potential benefit conferred by the BRAVO fast-path.  Furthermore, 
measuring the revocation duration also incorporates the waiting time, as well as the scanning time, 
yielding a conservative over-estimate of the revocation scan cost and resulting in less 
aggressive use of reader bias.
Despite these concerns, we find this policy yields good and predictable performance.   
For all benchmarks in this paper we used $N=9$ yielding a worst-case writer slow-down bound of about 10\%.  
Our policy required adding a second BRAVO-specific timestamp field \texttt{InhibitUntil} (Line~3), which 
reflects the earliest time at which slow readers should reenable bias\footnote{
We observe that it is trivial to collapse \texttt{RBias} and \texttt{InhibitUntil}
into just a single field.  For clarity, we did not do so in our implementation.}. 
We note that for safety, readers can only set \texttt{RBias} while they hold
read permission on the underlying reader-writer lock, avoiding interactions with writers (cf.~Lines~25--26).

In our implementations revoking waiters busy-wait for readers to depart.
There can be at most one such busy-waiting thread for a given lock at any given time.
We note, however, that it is trivial to shift to a waiting policy that uses blocking.



BRAVO acts as an accelerator layer, as readers can always fall back to the traditional underlying
lock to gain read access.  The benefit arises from avoiding coherence traffic on the centralized 
reader indicators in the underlying lock, and instead relying on updates to be diffused over the 
visible readers table.  Fast-path readers write only into the visible readers table, and not the
lock instance proper. 
This access pattern improves performance on NUMA systems, where write sharing is particularly expensive.
We note that if the underlying lock algorithm \emph{A} has reader preference or writer preference,
then \emph{BRAVO-A} will exhibit that same property.  
Write performance and the scalability of read-vs-write and write-vs-write behavior
depends solely on the underlying lock.  Under high write intensity, with write-vs-write 
and write-vs-read conflicts, the performance of BRAVO devolves to that of the underlying lock.
BRAVO accelerates reads only. 
BRAVO fully supports the case where a thread holds multiple locks at the same time.
\Invisible{Increases table occupancy and odds of collision, but rare in practice.
Holding multiple locks is equivalent to increasing concurrency with respect to table occupancy
and collision rates.} 


BRAVO supports \emph{try-lock} operations as follows.  For read \emph{try-lock} attempts
an implementation could try the BRAVO fast path and then fall back, if the fast path fails,
to the slow path underlying \emph{try-lock}.  An implementation can also opt to forgo the fast
path attempt and simply call the underlying \emph{try-lock} operator.
We use the former approach when applying BRAVO in the Linux kernel as detailed in the next section.
We note that if the underlying \emph{try-lock} call is successful, one may set \texttt{RBias}
if the BRAVO policy allows that (e.g., if the current time is larger than \texttt{InhibitUntil}).
For write \emph{try-lock} operators, an implementation will invoke the underlying
\emph{try-lock} operation.  If successful, and bias is set, then revocation must be performed
following the same procedure described in Lines~37--49.

As seen in Listing~\ref{Listing:BRAVO-py}, the \emph{slot} value must be passed from 
the read lock operator to the corresponding unlock. \texttt{Null} indicates that the slow
path was used to acquire read permission.
To provide correct \texttt{errno} error return values in the POSIX pthread environment, a thread must be able
to determine if it holds read, write, or no permission on a given lock.  This is
typically accomplished by using per-thread lists of locks currently held in read mode.
We leverage those list elements to pass the slot.  We note that the Cohort read-write lock
implementation~\cite{ppopp13-calciu} passed the reader's NUMA node ID from lock to corresponding unlock in
this exact fashion.

\section{Applying BRAVO to the Linux kernel \texttt{rwsem}}
\label{sec:rwsem} 
In this section, we describe prototype integration of BRAVO in the Linux kernel, where we apply it to \texttt{rwsem}.
\texttt{Rwsem} is a read-write semaphore construct.
Among many places inside the kernel, it is used to protect the access to the virtual memory area (VMA) structure of 
each process~\cite{Cor18}, which makes it a source of contention for data intensive applications~\cite{Cor18, CKZ12}.

On a high level, \texttt{rwsem} consists of a counter and a waiting queue protected by a spin-lock.
The counter keeps track of the number of active readers, as well as encodes the presence of a writer.
To acquire the \texttt{rwsem} in the read mode, a reader atomically increments the counter and checks its value.
If a (waiting or active) writer is not present, the read acquisition is successful; otherwise, the reader acquires the 
spin-lock protecting the waiting queue, joins the queue at the tail, releases the spin-lock and blocks, 
waiting for a wake-up signal from a writer.
As a result, when there is no reader-writer contention, the read acquisition boils down to 
one atomic counter increment.
On architectures that do not support an atomic increment instruction, 
this requires acquisition (and subsequent release) of the spin-lock.
Even on architectures that have such an instruction (such as Intel x86), the read acquisition of  
\texttt{rwsem} creates contention over the cache line hosting the counter.

In our integration of BRAVO on top of \texttt{rwsem}, we make a simplifying assumption that the semaphore 
is always released by the same thread that acquired it for read.
This is not guaranteed by the API of \texttt{rwsem}, however, this is a common way of using semaphores 
in the kernel.
This assumption allows us to preserve the existing \texttt{rwsem} API and limits the scope of changes required, resulting in a patch
of only three files and adding just a few dozens lines of code.
We use this assumption when determining the slot into which a thread would store the semaphore address 
on the fast acquisition path, and clear that slot during the release operation\footnote{We determine the slot 
by hashing the task struct pointer (\texttt{current}) with the address of the semaphore.}.

While we have not observed any issue when running and evaluating the modified kernel, we note that our assumption
can be easily eliminated by, for example, extending the API of \texttt{rwsem} to allow an additional pointer argument for read acquisition and
release functions.
In case the acquisition is made on the fast path, this pointer would be used to store the address of the corresponding slot; 
later, this pointer can be passed to a (different) releasing thread to specify the slot to be cleared.
Alternatively, we can extend the API of \texttt{rwsem} to include a flag explicitly allowing the use of the fast path for read acquisition and release.
This flag would be set only in call sites known for high read contention (such as in functions operating on VMAs), where a
thread that releases the semaphore is known to be the one that acquired it.
Other call sites for semaphore acquisition and release can be left untouched, letting them use the slow path only.

We note that the default configuration of the kernel enables a so-called spin-on-owner optimization of \texttt{rwsem}~\cite{Long17}. 
With this optimization, the \texttt{rwsem} structure includes an \texttt{owner} field that contains a pointer to the \texttt{current} struct
of the owner task when \texttt{rwsem} is acquired for write.
Using this field, a reader may check whether the writer is currently running on a CPU, and if so, spin rather than block~\cite{Long17}.
While writers do not use this field to decide whether they have to spin (as there might be multiple readers), in the current \texttt{rwsem} implementation
a reader updates the \texttt{owner} field regardless, storing there its \texttt{current} pointer 
along with a few control bits (that specify that the lock is owned by a reader).
These writes by readers are for debugging purposes only, yet they create unnecessary contention on the \texttt{owner} field.
We fix that by letting a reader set only the control bits in the \texttt{owner} field, 
and only if those bits were not set before, i.e., when the first reader acquires that \texttt{rwsem} instance after a writer.
Note that all subsequent readers would read, but not update the \texttt{owner} field, until it is updated again by a writer.

\section{User-space Evaluation}

All user-space data was collected on an Oracle X5-2 system.  The system has 2 sockets, each populated with 
an Intel Xeon E5-2699 v3 CPU running at 2.30GHz.  Each socket has 18 cores, and each core is 2-way 
hyperthreaded, yielding 72 logical CPUs in total.  The system was running Ubuntu 18.04 with a stock 
Linux version 4.15 kernel, and all software was compiled using the provided GCC version 7.3 toolchain
at optimization level ``-O3''.  
64-bit code was used for all experiments.  Factory-provided system defaults were used in all cases, 
and \emph{Turbo mode}~\cite{turbo} was left enabled.  In all cases default free-range unbound threads were used.  

We implemented all locks within LD\_PRELOAD interposition
libraries that expose the standard POSIX \texttt{pthread\_\allowbreak{}rwlock\_t} programming interface. 
This allows us to change lock implementations by varying the LD\_PRELOAD environment variable and
without modifying the application code that uses reader-writer locks.
This same framework was used to implement Cohort reader-writer locks \cite{ppopp13-calciu}.

In the following figures ``BA'' refers to the Brandenburg-Anderson PF-Q lock~\cite{rtsj10-brandenburg}; ``Cohort-RW'' refers to the
C-RW-WP lock~\cite{ppopp13-calciu}; ``Per-CPU'' reflects a lock that consists of an array of
BA locks, one for each CPU, where readers acquire read-permission on the sub-lock associated with their 
CPU, and writers acquire write-permission on all the sub-locks (this lock is inspired by the Linux kernel 
brlock construct~\cite{brlock2}); ``BRAVO-BA'' reflects BRAVO implemented
on top of BA; ``pthread'' is the default Linux POSIX ``pthread\_rwlock'' read-write lock mechanism; and
``BRAVO-pthread'' is BRAVO implemented on top of the pthread\_rwlock.  

\Invisible{Arrival cost vs waiting cost} 

We also took data on the Brandenburg-Anderson PF-T lock and the BRAVO form thereof.  
PF-T implements the reader indicator via a central pair of counters, one incremented by arriving readers
and the other incremented by departing readers.
Waiting readers busy-wait on a dedicated \emph{writer present} bit encoded in the reader arrival counter.  
In PF-Q active readers are tallied on a central pair of counters in the same fashion as PF-T,
but waiting readers enqueue on an MCS-like queue.
In both PF-T and PF-Q, arriving readers update the central reader indicator state, generating more coherence traffic
than would be the case for locks that use distributed reader indicators or BRAVO.  
Waiting (blocked) readers in PF-T use global spinning, while waiting readers in PF-Q use local spinning
on a thread-local field in the enqueued element.  
PF-T enjoys slightly shorter code paths but also suffers from lessened scalability because of the global spinning.
We found that PF-T and PF-Q offer broadly similar performance, with PF-T having a slight advantage 
when the arrival rate is high, the number of waiting threads is low, and the waiting period is shorter.   
PF-T is slightly more compact having just 4 integer fields, while PF-Q has 2 such fields and 4 pointers.   
For brevity, we do not include PF-T results.  
We also found that ``fair lock with local only spinning'' by Mellor-Crummey and Scott~\cite{ppopp91-Mellor-Crummey} 
yielded performance similar to or slower than that of PF-Q.  

We note that the default pthread read-write lock implementation found in our Linux distribution provides
strong reader preference, and admits indefinite writer starvation
\footnote{The pthread implementation allows writer preference to be selected via a non-portable API.  
Unfortunately in the Linux distribution we used this feature has bugs that result in lost wakeups and hangs: 
\url{https://sourceware.org/bugzilla/show_bug.cgi?id=23861}.}. 
The reader indicator is centralized and the lock has a footprint of 56 bytes 
for 64-bit programs.  Waiting threads block immediately in the kernel without spinning.  While this policy incurs
overheads associated with voluntary context switching, it may also yield benefits by allowing ``polite'' waiting 
by enabling Turbo mode for those threads making progress.  
Except where otherwise noted, we plot the number of concurrent threads on the X-axis, and 
aggregate throughput on the Y-axis reporting the median of 7 independent runs for each data point.



We use a 128 byte sector size on Intel processors for alignment to avoid 
false sharing.  The unit of coherence is 64 bytes throughout the cache hierarchy, but 128 bytes
is required because of the adjacent cache line prefetch facility where pairs of lines are automatically 
fetched together.  BA requires just 128 bytes -- 2 32-bit integer fields plus 4 pointers fields with the overall size rounded up 
to the next sector boundary.   BRAVO-BA adds the 8-byte \texttt{InhibitUntil} field, which contains 
a timestamp, and the 4-byte \texttt{RBias} field.  Rounding up to the sector size, this still yields 
a 128 byte lock instance.  Per-CPU consists of one instance of BA for each logical CPU, yielding 
a lock size of 9216 bytes on our 72-way system.  Cohort-RW consists of one reader indicator 
(128 bytes) per NUMA node, a central location for state (128 bytes) and a full cohort mutex 
\cite{topc15-dice} to provide writer exclusion.  In turn, the cohort mutex requires one
128-byte sub-lock per NUMA node, and another 256 bytes for central state, for a total of 896 bytes.  
(While our implementation did not do so, we note that a more space aggressive implementation of Cohort-RW
could colocate the per-node reader indicators with the mutex sub-locks, and the central state for the
read-write lock with its associated cohort mutex, yielding a size of 512 bytes).  
As noted above, the size of the pthread read-write lock is 56 bytes, and the 
BRAVO variant adds 12 bytes.  The size of BA, BRAVO-BA, pthread, and BRAVO-pthread are fixed, and known at 
compile-time, while the size of Per-CPU varies with the number
of logical CPUs, and the size of Cohort-RW varies with the number of NUMA nodes.  
Finally, we observe that BRAVO allows more relaxed approach toward the alignment and padding of
the underlying lock.  Since fast-path readers do not mutate the underlying lock fields, the
designer can reasonable forgo alignment and padding on that lock, without trading off
reader scalability.  

The size of the lock can be important in concurrent data structures, such as linked lists or
binary search trees, that use a lock per node or entry \cite{ppopp10-Bronson, ppopp12-Crain, opodis05-Heller}.
As Bronson at el. observe, when a scalable lock is striped across multiple cache lines to
avoid contention in the coherence fabric, it is ``prohibitively
expensive to store a separate lock per node''\cite{ppopp10-Bronson}.

BRAVO also requires the visible readers table.  With 4096 entries on a system with 64-bit pointers,
the additional footprint is 32KB.  The table is aligned and sized to minimize the number of underlying
pages (reducing TLB footprint) and to eliminate false sharing from variables that might be placed
adjacent to the table.  We selected a table size 4096 empirically but in general believe the size
should be a function of the number of logical CPUs in the system.  Similar tables in the Linux kernel,
such as the \emph{futex} hash table, are sized in this fashion~\cite{futexes}. 
\footnote{\url{https://blog.stgolabs.net/2014/01/futexes-and-hash-table-collisions.html}}.

\Invisible{Ideally we'd place the visible readers array on large pages to reduce TLB pressure.}  

BRAVO yields a favorable performance trade-off between space and scalability,
offering a viable alternative that resides on the design spectrum between classic centralized locks, such
as BA, having small footprint and poor reader scalability, and the large locks with high reader scalability.

\subsection{Sensitivity to Inter-Lock Interference} 

\begin{figure}[h]                                                                    
\includegraphics[width=8.5cm]{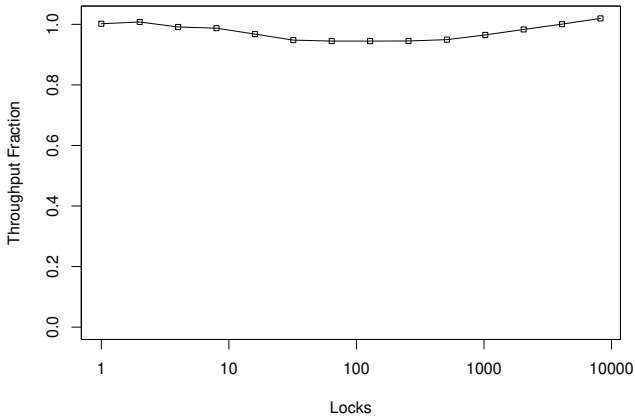}
\caption{Inter-Lock Interference}                                                   
\label{Figure:Interference}                                                                  
\end{figure}

As the visible readers array is shared over all locks and threads within an address space, one potential
concern is collisions and near collisions that might arise when multiple threads are using a large set of
locks.  Near collisions are also of concern as they can cause false sharing within the array.   To determine BRAVO's
performance sensitivity to such effects, we implemented a microbenchmark program that spawns 64 concurrent threads. 
Each thread loops as follows: randomly pick a reader-writer lock from a pool of such locks; acquire that lock for
read; advance a thread-local pseudo-random number generator 20 steps; release read permission on the lock; and finally
advance that random number generator 100 steps.  At the end of a 10 second measurement interval we report the number
of lock acquisitions.  No locks are ever acquired with write permission.   Each data point is the median of 7 distinct runs.
We report the results in Figure~\ref{Figure:Interference} where the X-axis reflects the number of locks in the pool (varying through
powers-of-two between 1 and 8192) and the Y-axis is the number of acquisitions completed by BRAVO-BA divided 
by the number completed by a specialized version of BRAVO-BA where each lock instance has a private array of 4096 elements.  
This fraction reflects the performance drop attributable to inter-lock conflicts and near conflicts in the shared array, where 
the modified form of BRAVO-BA can be seen as an idealized form that has a large per-instance footprint but which is
immune to  inter-lock conflicts\footnote{We note that as we increase the number of locks, cache pressure constitutes 
a confounding factor for the specialized version of BRAVO-BA.}. 
The worst-case penalty arising from inter-thread interference (the lowest fraction value) 
is always under 6\%. 

\Invisible{See InterferenceRW.cc.  
There may exists better experiments to show that conflicts aren’t important -- insensitivity.
For instance, we could pick some fixed configuration of locks and threads and vary the BRAVO array size, 
showing that the function from size to performance is asymptotic and that our choice of size is selected 
so that interference is negligible and positioned "far out" on the long tail of the asymptote.} 
     
\subsection{Alternator}

\begin{figure}[tp]                                                                    
\includegraphics[width=8.5cm]{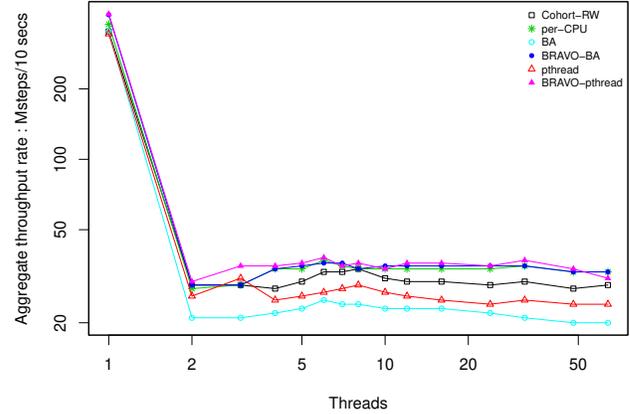}
\caption{Alternator}                                                   
\label{Figure:Alternator}                                                                  
\end{figure}

Figure~\ref{Figure:Alternator} shows the results of our \texttt{alternator} benchmark.  
The benchmark spawns the specified number of concurrent threads, which organize themselves into a logical 
ring, each waiting for notification 
from its ``left'' sibling.   Notification is accomplished via setting a thread-specific variable via
a store instruction,
and waiting is via simple busy-waiting.  Once notified, the thread acquires and then immediately releases 
read permission on a shared reader-writer lock.  Next the thread notifies its ``right'' sibling and 
then again waits.  There are no writers, and there is no concurrency between readers.  
At most one reader is active at any given moment.
At the end of a 10 second measurement interval the program reports the number of notifications. 

The BA lock suffers as the lines underlying the reader indicators ``slosh'' and migrate from cache 
to cache.  In contrast BRAVO-BA readers touch different locations in the visible readers table 
as they acquire and release read permissions.  BRAVO enables reader-bias early in the run, and it remains
set for the duration of the measurement interval.  
All locks experience a significant performance drop between 1 and 2 threads due to the impact 
of coherent communication for notification.  Crucially, we see that BRAVO-BA outperforms the 
underlying BA by a wide margin, and is competitive with the much larger Per-CPU lock. 
In addition, the performance of BA can be seen to degrade as we add threads, whereas the
performance of BRAVO-BA remains stable.
The same observations are true when considering BRAVO-pthread and pthread locks.

Since the hash function that associates a read locking request with an index is deterministic,
threads repeatedly locking and unlocking a specific lock will enjoy temporal locality and reuse
in the visible readers table.  

\subsection{test\_rwlock}

\begin{figure}[tp]                                                                    
\includegraphics[width=8.5cm]{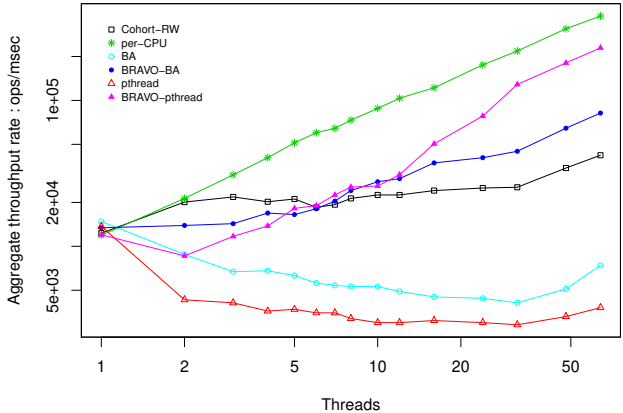}
\caption{test\_rwlock}                                                   
\label{Figure:rwlock}                                                                  
\end{figure}

We next report results from the \texttt{test\_rwlock} benchmark described by Desnoyers et al.~\cite{tpds12-desnoyers}%
\footnote{obtained from \url{https://github.com/urcu/userspace-rcu/blob/master/tests/benchmark/test_rwlock.c} 
and modified slightly to allow a fixed measurement interval.}. 
The benchmark was designed to evaluate the performance and scalability of reader-writer locks against 
the RCU (Read-Copy Update) synchronization mechanism.  We used the following command-line:
\texttt{test\_rwlock T 1 10 -c 10 -e 10 -d 1000}.
The benchmark launches 1 fixed-role writer thread and $T$ fixed-role reader threads for
a 10 second measurement interval.  The writer loops as follows : acquire a central reader-writer lock instance; 
execute 10 units of work, which entails counting down a local variable; release writer permission; execute a non-critical
section for 1000 work units.  Readers loop acquiring the central lock for reading, executing 10 steps of work in the 
critical section, and then release the lock.  (The benchmark has no facilities to allow a non-trivial critical
section for readers).  At the end of the measurement interval the benchmark reports the sum of
iterations completed by all the threads. 
As we can see in Figure~\ref{Figure:rwlock}, BRAVO-BA significantly outperforms BA, and even the Cohort-RW lock
at higher thread counts.  Since the workload is extremely read-dominated, the Per-CPU lock yields
the best performance, albeit with a very large footprint and only because of the relatively low write rate.
For that same reason, and due to its default reader preference, BRAVO-pthread easily beats pthread, and comes close
to the performance level of Per-CPU.

\subsection{RWBench}
\begin{figure*}[hpt!] 
\centering
\begin{subfigure}{.48\textwidth}                                                                   
\centering
\includegraphics[width=8.5cm]{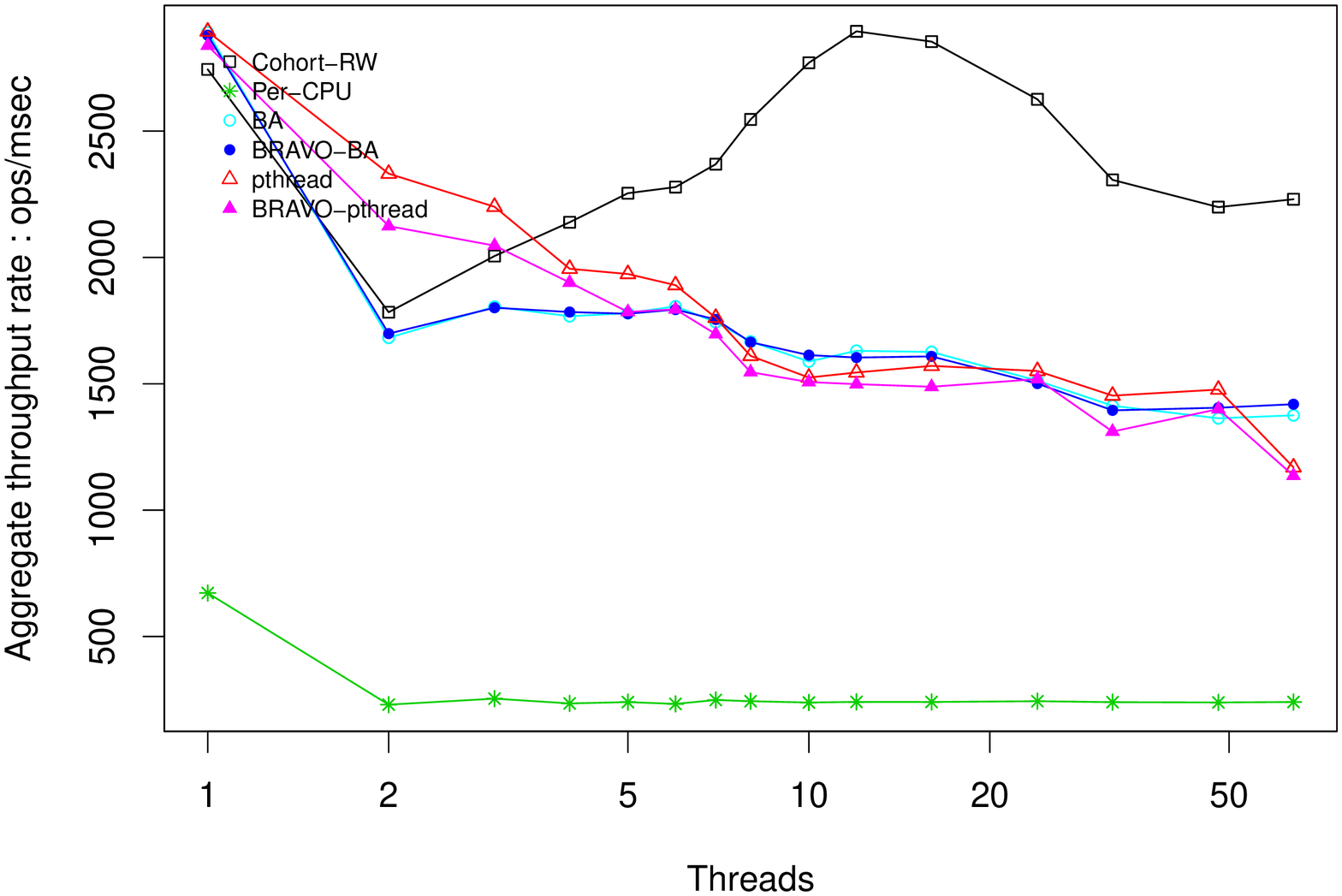}
\caption{RWBench with 90\% writes (9/10)}                                                   
\hfill
\label{Figure:RWBench-Point9}                                                                  
\end{subfigure}   
\begin{subfigure}{.48\textwidth}                                                      
\centering
\includegraphics[width=8.5cm]{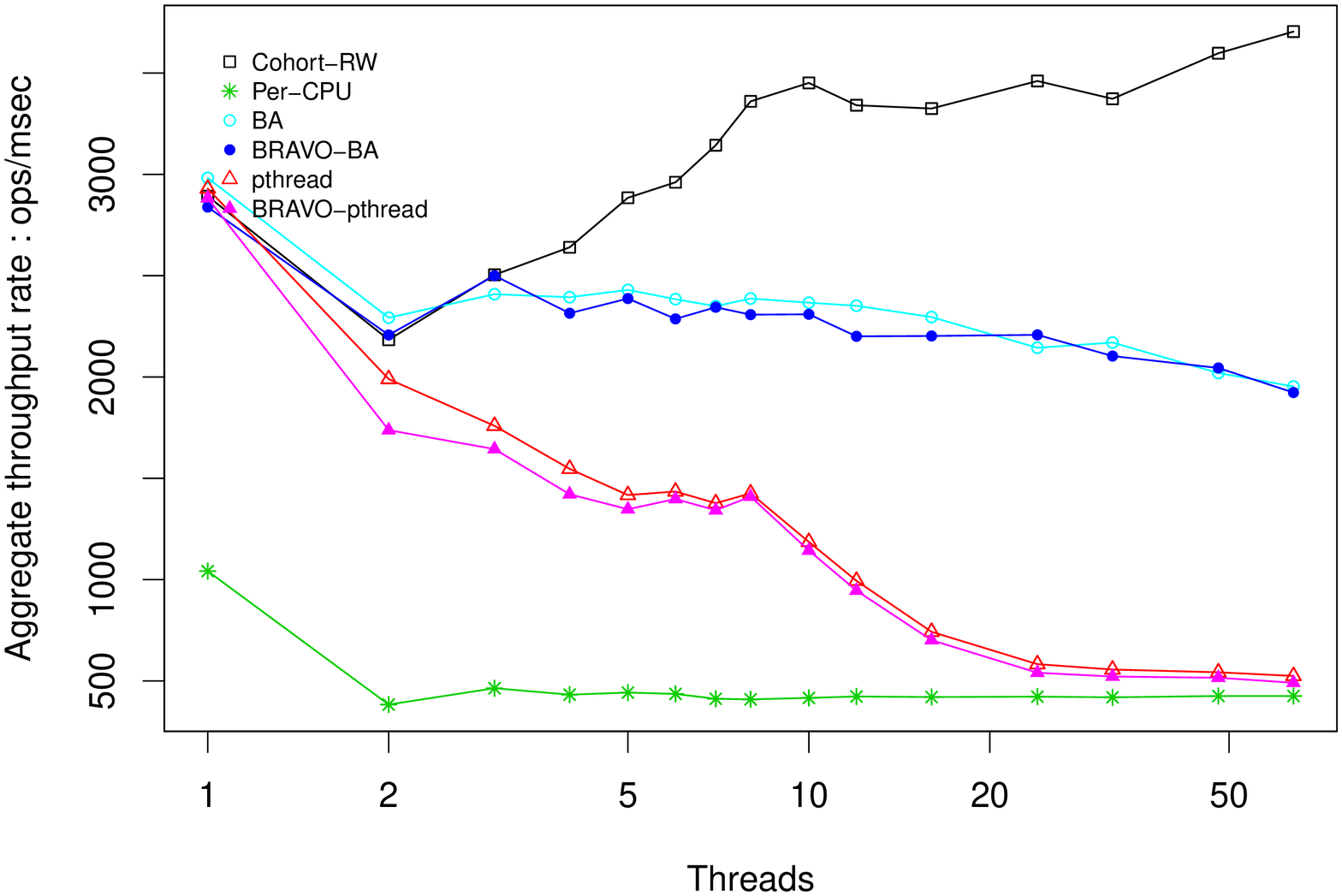}
\caption{RWBench with 50\% writes (1/2)}                                                   
\hfill
\label{Figure:RWBench-Point5}                                                                  
\end{subfigure}   
\begin{subfigure}{.48\textwidth}                                                               
\centering
\includegraphics[width=8.5cm]{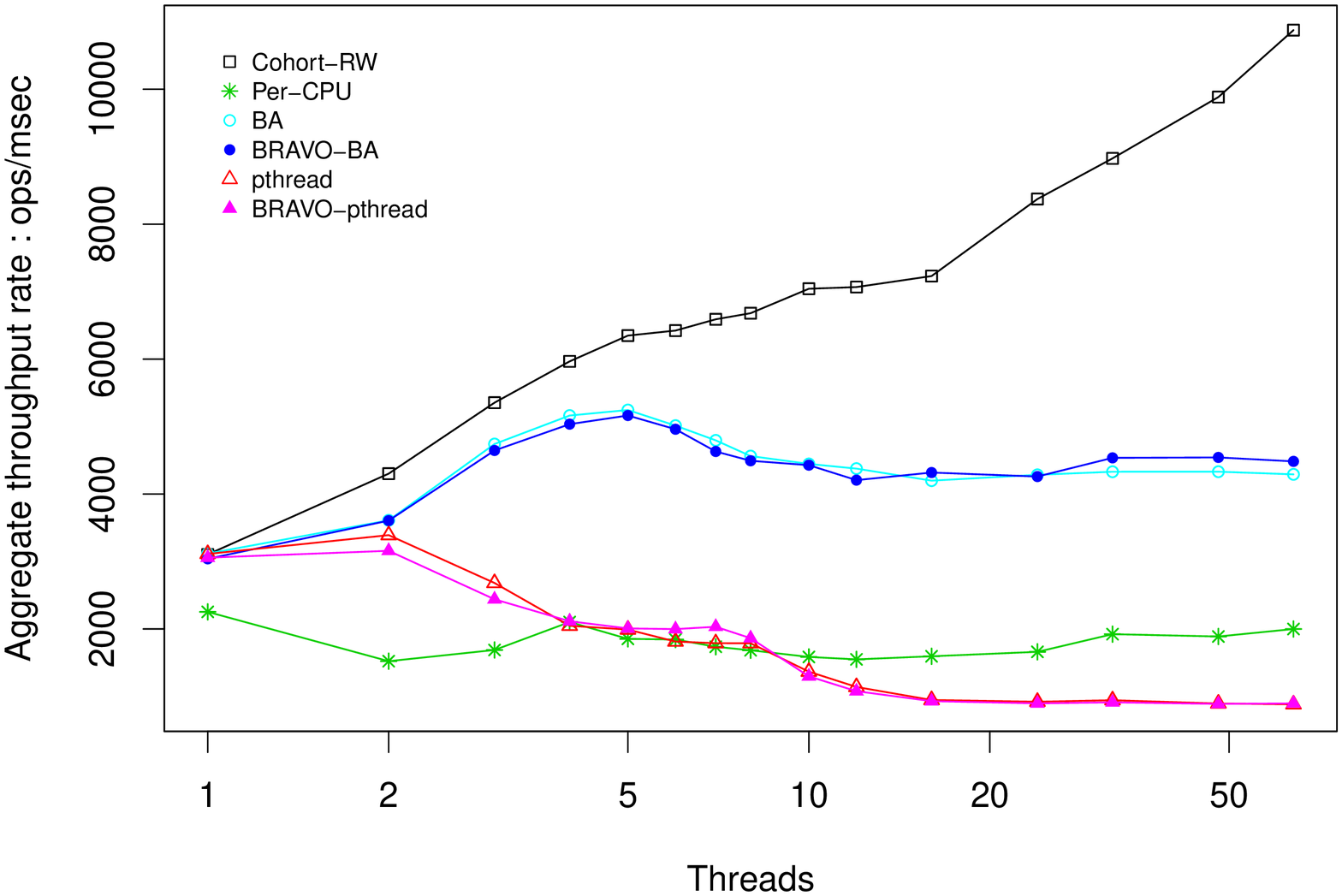}
\caption{RWBench with 10\% writes (1/10)}                                                   
\hfill
\label{Figure:RWBench-10}                                                                  
\end{subfigure}   
\begin{subfigure}{.48\textwidth}                                                              
\centering
\includegraphics[width=8.5cm]{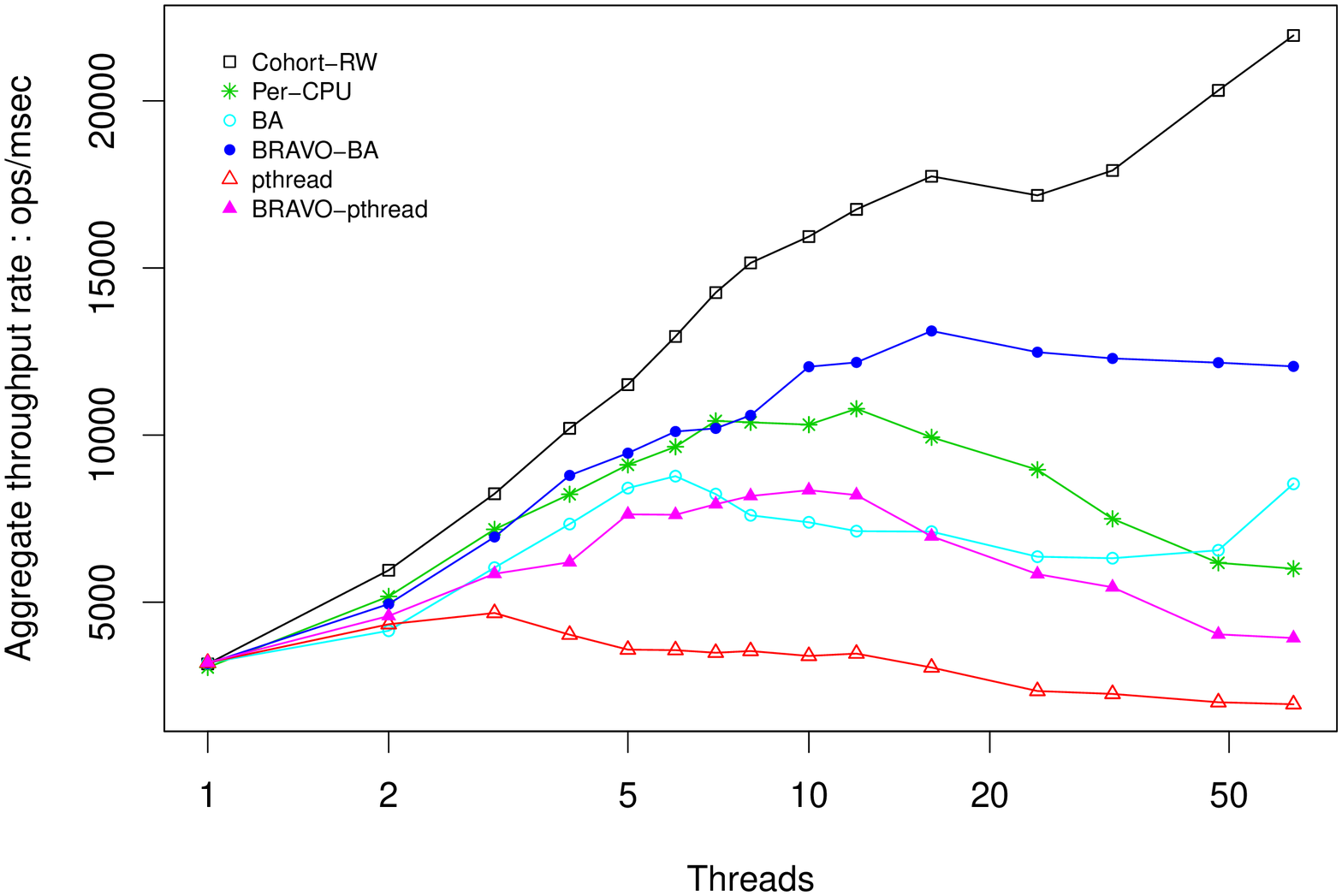}
\caption{RWBench with 1\% writes (1/100)}                                                   
\hfill
\label{Figure:RWBench-100}                                                                  
\end{subfigure}   
\begin{subfigure}{.48\textwidth}                                                                    
\centering
\includegraphics[width=8.5cm]{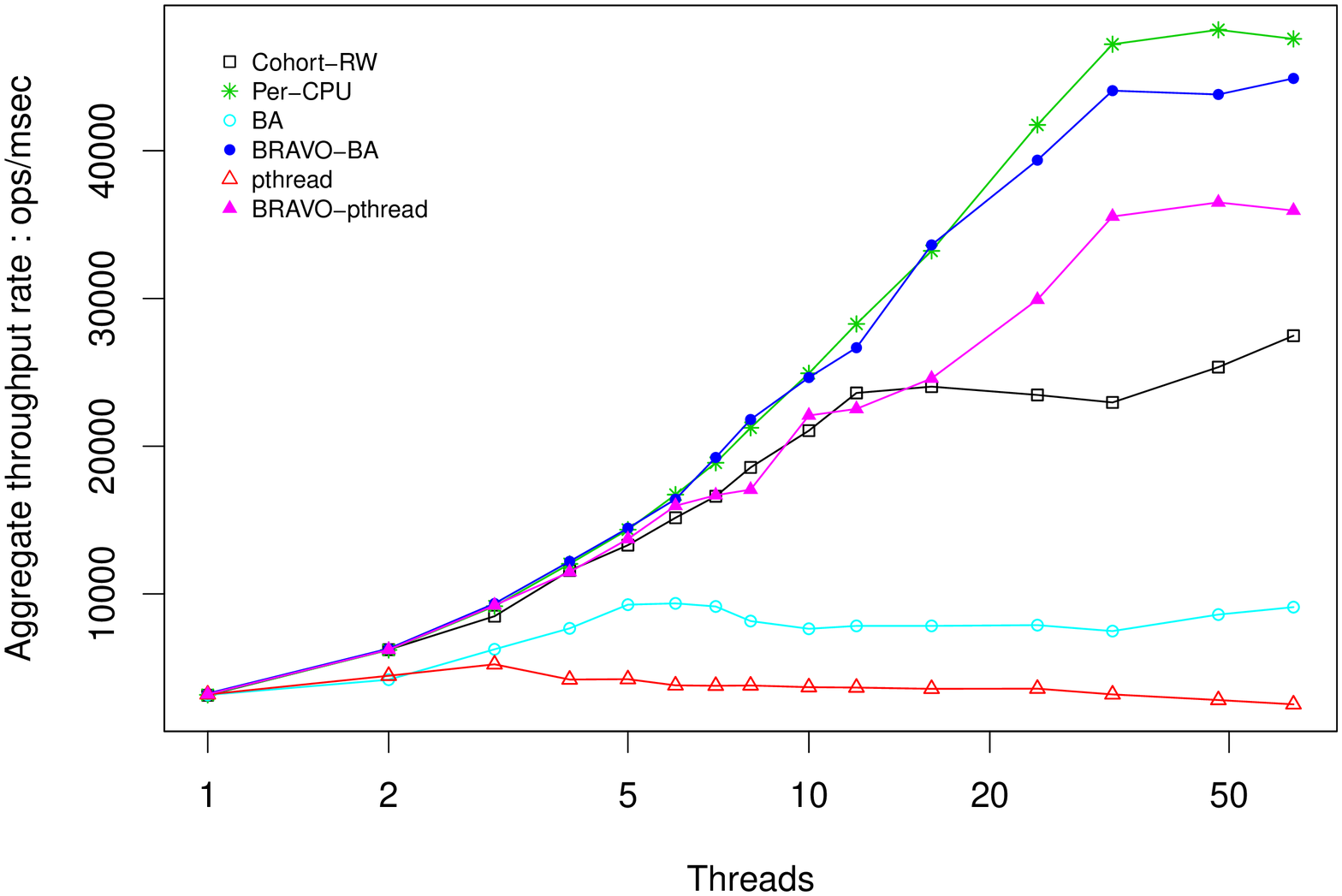}
\caption{RWBench with .1\% writes (1/1000)}                                                   
\hfill
\label{Figure:RWBench-1000} 
\end{subfigure}   
\begin{subfigure}{.48\textwidth}                                                                
\centering
\includegraphics[width=8.5cm]{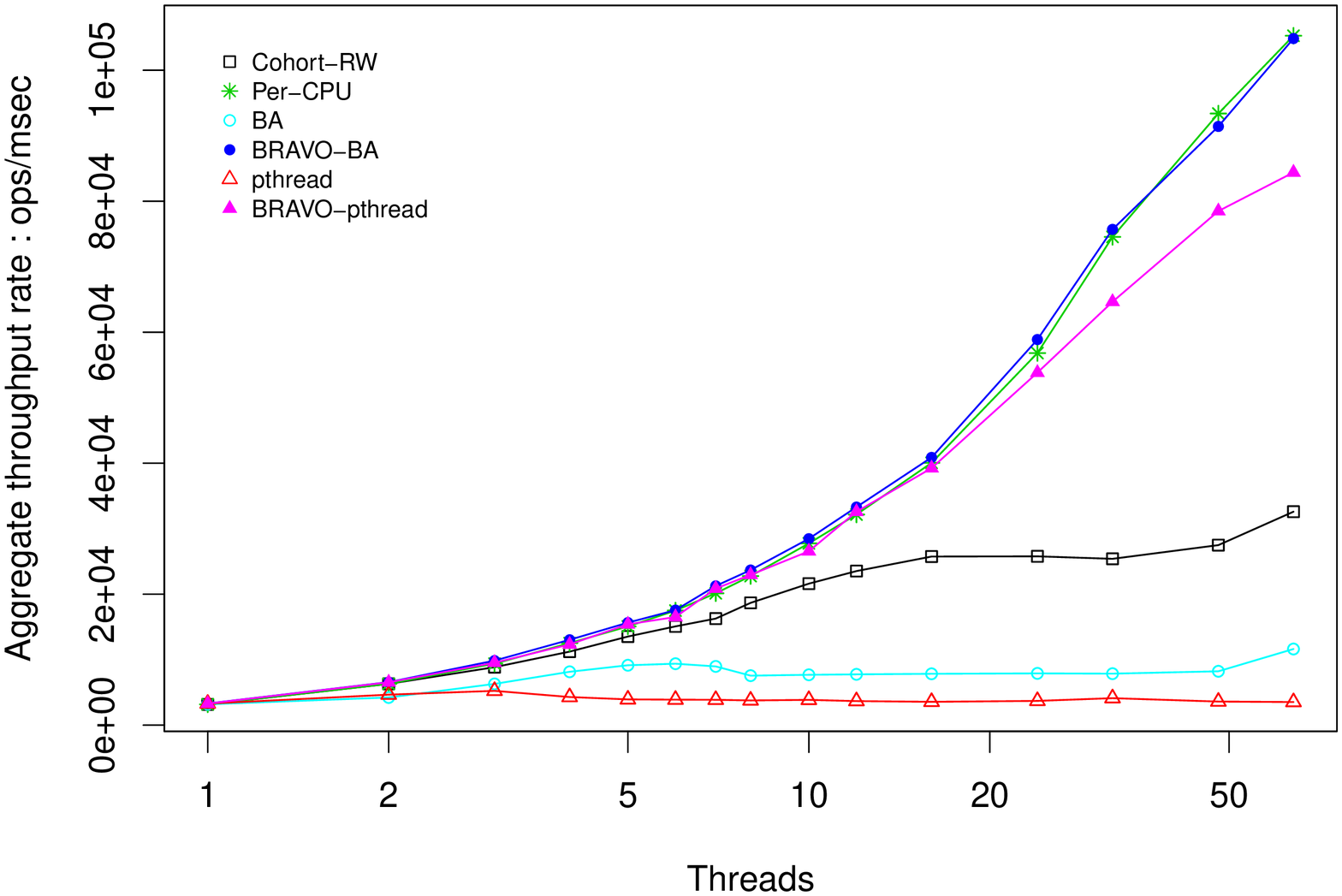}
\caption{RWBench with .01\% writes (1/10000)}                                                   
\hfill
\label{Figure:RWBench-10000}                                                                  
\end{subfigure}   
\caption{RWBench} 
\end{figure*}

Using \texttt{RWBench} -- modeled on a benchmark of the same name described by Calciu et al.~\cite{ppopp13-calciu} 
-- we evaluated the reader-write lock algorithms over a variety of read-write ratios, ranging 
from write-intensive in Figure~\ref{Figure:RWBench-Point9} (9 out of every 10 operations are writes) 
to read-intensive in Figure~\ref{Figure:RWBench-10000} (1 out of every 10000 operations are writes), 
demonstrating that BRAVO inflicts no harm for write-intensive workloads, but improves performance
for more read-dominated workloads.  \texttt{RWBench} launches $T$ concurrent threads for a 10 second 
measurement interval.
Each thread loops as follows : using a thread-local pseudo-random generator, decide
to write with probability $P$ via a Bernoulli trial; writers acquire a central reader-write lock
for write permission and then execute 10 steps of a thread-local C++ \texttt{std::mt19937} 
random number generator and then release write permission, while readers do the same, but under
read permission; execute a non-critical section of $N$ steps of the same random-number generator
where $N$ is a random number uniformly distributed in $[0,200)$ with average and median of 100.  
At the end of the measurement interval the benchmark reports the total number of top-level loops 
completed.  

In Figure~\ref{Figure:RWBench-Point9} we see poor scalability over all the locks by virtue of the
highly serialized write-heavy nature of the workload.  Per-CPU fairs poorly as writes, which are common,
need to scan the array of per-CPU sub-locks.  Cohort-RW provides some benefit, while
BRAVO-BA (BRAVO-pthread) tracks closely to BA (pthread, respectively), providing neither benefit nor harm.  
The same behavior plays out in Figure~\ref{Figure:RWBench-Point5} ($P = 1/2$) and Figure~\ref{Figure:RWBench-10} 
($P = 1/10$), although in the latter we see some scaling from Cohort-RW.  
In Figure~\ref{Figure:RWBench-100} ($P = 1/100$) we begin to see BRAVO-BA outperforming BA at higher thread
counts.  Figure~\ref{Figure:RWBench-1000} ($P = 1/1000$) and Figure~\ref{Figure:RWBench-10000} 
($P = 1/10000$ -- extremely read-dominated) are fairly similar, with BRAVO-BA and BRAVO-pthread yielding
performance similar to that of Per-CPU, Cohort-RW yielding modest scalability,
and BA and pthread yielding flat performance as the thread count increases.


\subsection{\texttt{rocksdb} readwhilewriting} 

\begin{figure}[h]                                                                    
\includegraphics[width=8.5cm]{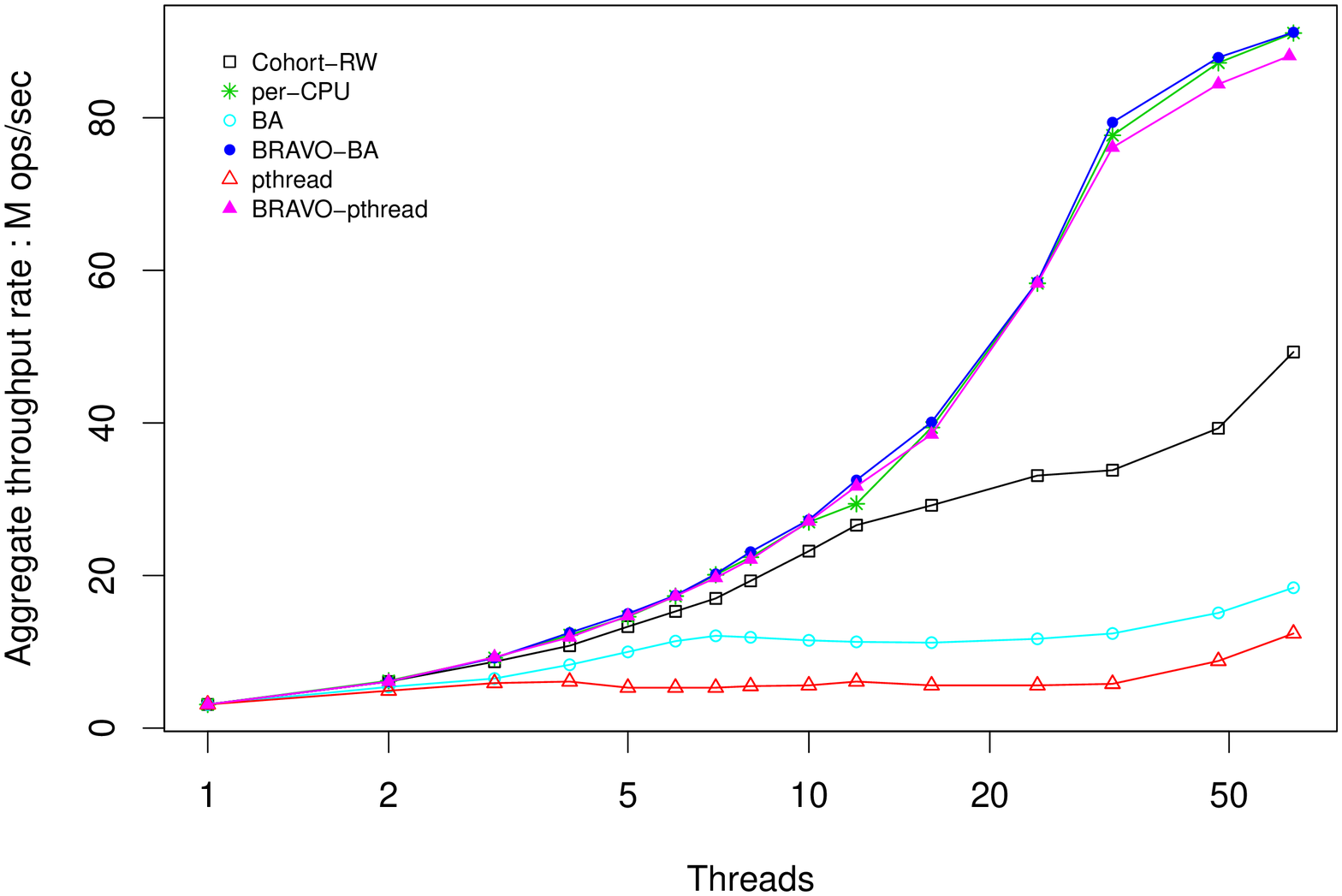}
\caption{\texttt{rocksdb} readwhilewriting} 
\label{Figure:rocksdb-readwhilewriting}                                                                  
\end{figure}

We next explore performance sensitivity to reader-writer in the rocksdb database~\cite{rocksdb}.  
We observed high frequency reader traffic arising from calls in \texttt{::Get()} to 
\texttt{db/memtable.cc GetLock()} in the \texttt{readwhilewriting} benchmark
\footnote{We used \texttt{rocksdb} version 5.13.4 with the following command line: \texttt{db\_bench --threads=T --benchmarks=readwhilewriting --memtablerep=cuckoo  -duration=100 --inplace\_update\_support=1 
--allow\_concurrent\_memtable\_write=0 --num=10000 --inplace\_update\_num\_locks=1 
--histogram --stats\_interval=10000000}}.  In Figure~\ref{Figure:rocksdb-readwhilewriting} we see the performance 
of BRAVO-BA and BRAVO-pthread tracks that of Per-CPU and always exceeds that of Cohort-RW and the respective underlying locks.

\subsection{\texttt{rocksdb} hash\_table\_bench}

\texttt{Rocksdb} also provides a benchmark to stress the hash table used by their persistent cache
\footnote{\url{https://github.com/facebook/rocksdb/blob/master/utilities/persistent_cache/hash_table_bench.cc} 
run with the following command-line: \texttt{hash\_table\_bench -nread\_thread=$T$ -nsec=50}}. 
The benchmark implements a central shared hash table as a C++ \texttt{std::unordered\_map} protected by a 
reader-writer lock.  The cache is pre-populated before the measurement interval.  At the end of 
the 50 second measurement interval the benchmark reports the aggregate operation rate -- reads, 
erases, insertions -- per millisecond.  A single dedicated thread loops, erasing random elements, 
and another dedicated thread loops inserting new elements with a random key.  Both erase and insertion
operations require write access.  The benchmark launches 
$T$ reader threads, which loop, running lookups on randomly selected keys.  We vary $T$ on the X-axis. 
All the threads execute operations back-to-back without a delay between operations. 
The benchmark makes frequent use of malloc-free operations in the \texttt{std::unordered\_map}.  
The default malloc allocator fails to fully scale in this environment and masks any benefit conferred by
improved reader-writer locks, so we instead used the index-aware allocator by Afek et al.~\cite{ismm11-afek}.

\begin{figure}[h]                                                                    
\includegraphics[width=8.5cm]{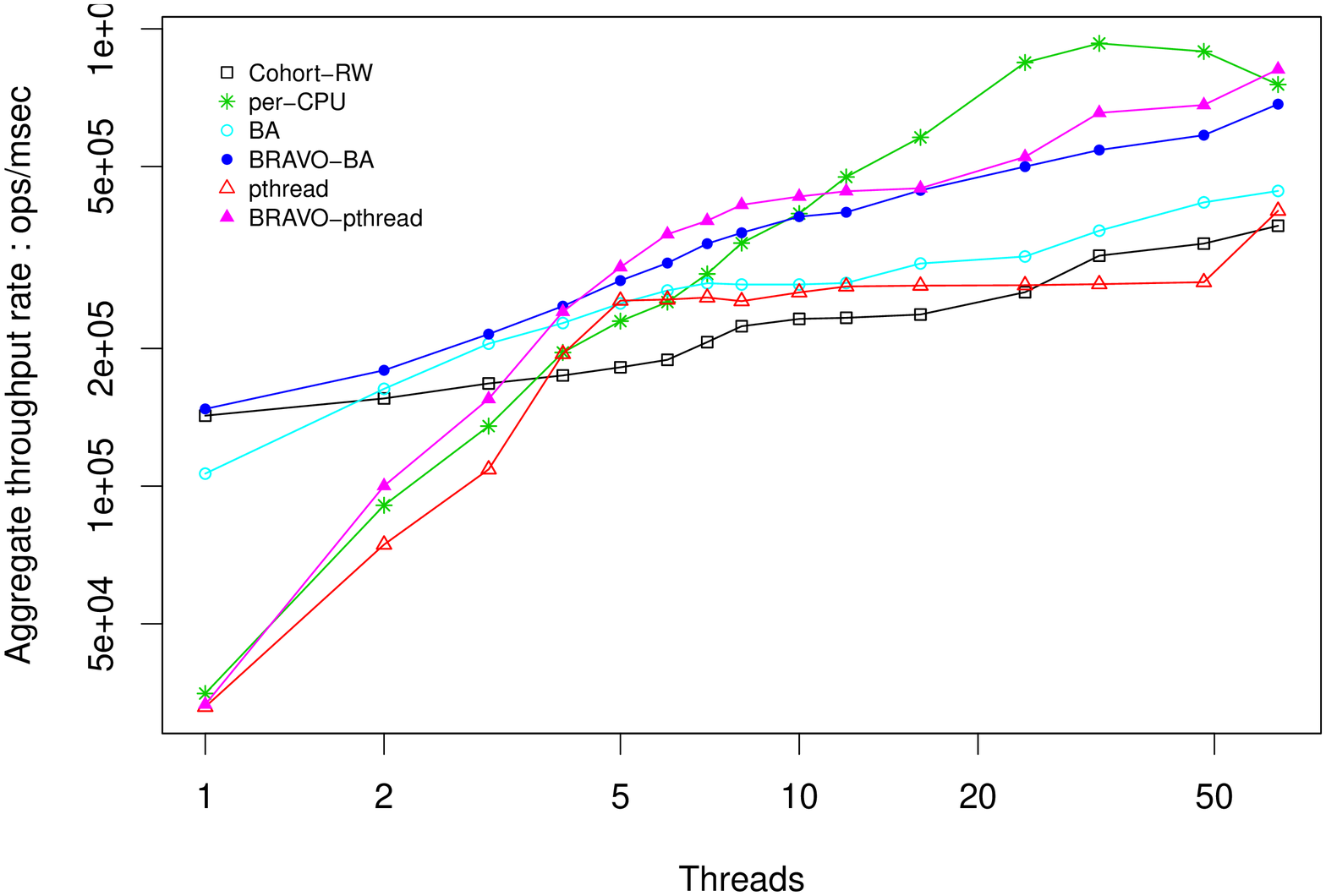}
\caption{\texttt{rocksdbi} hash\_table\_bench with \texttt{std::unordered\_map}} 
\label{Figure:rocksdb-hashtablebench}                                                                  
\end{figure}

The results are shown in Figure~\ref{Figure:rocksdb-hashtablebench}.
Once again, BRAVO enhances the performance of underlying locks, and shows substantial speedup
at high thread counts.

\section{Linux Kernel Experiments}
\label{sec:kernel-evaluation} 
All kernel-space data was collected on an Oracle X5-4 system.  The system has 4 sockets, each populated with 
an Intel Xeon CPU E7-8895 v3 running at 2.60GHz.  Each socket has 18 cores, and each core is 2-way 
hyperthreaded, yielding 144 logical CPUs in total.  The patch was applied on top of a recent
Linux version 4.20.rc4 kernel, which refer to as \emph{stock}. 
Factory-provided system defaults were used in all cases.
In particular, we compiled the kernel in the default configuration, which notably disables the lock performance data collection
mechanism (aka \texttt{lockstat}\footnote{\url{https://www.kernel.org/doc/Documentation/locking/lockstat.txt}}) built into the kernel
for debugging lock performance.
As we mention below, this mechanism was useful to gain insights into the usage patterns of kernel locks under various applications.
However, we kept it disabled during performance measurements as it adds a probing effect by generating stores into shared variables, 
e.g., by keeping track of the last CPU on which a given lock instance, \texttt{rwsem} included, was acquired.
These stores hamper the benefit of techniques like BRAVO that aim to reduce 
updates to the shared state during lock acquisition and release operations.

Each experiment is repeated $7$ times, and the reported numbers are the average of the corresponding results.
Unless noted, the reported results were relatively stable, with variance of less than $5\%$ from the average in most cases.
In the following, we refer to the kernel version modified to use BRAVO simply as BRAVO.

\subsection{locktorture}
\texttt{locktorture} is a loadable kernel module distributed with the kernel. 
As its name suggests, \texttt{locktorture} contains a set of microbenchmarks for 
evaluating performance of various synchronization constructs in the kernel, including \texttt{rwsem}.
It allows specifying the number of readers and writers that repeatably acquire the \texttt{rwsem} in the corresponding mode, and 
hold it (aka run in a critical section) for some amount of time.
The typical length of the critical section is 50ms for readers and 10ms for writers.
Occasionally, a long delay is introduced (according to a comment in the source code, 
``to force massive contention''), which means a critical section of 200ms for a reader and of 1000ms for a writer.
We note that the probability for those delays is small, yet it depends on the number of threads.
Therefore, the average length of the critical section is not the same across all thread counts, and thus the reported results do
not necessarily measure scalability.

\begin{figure}

\begin{subfigure}{0.25\textwidth}                                                                   
\centering
\includegraphics[width=1\textwidth]{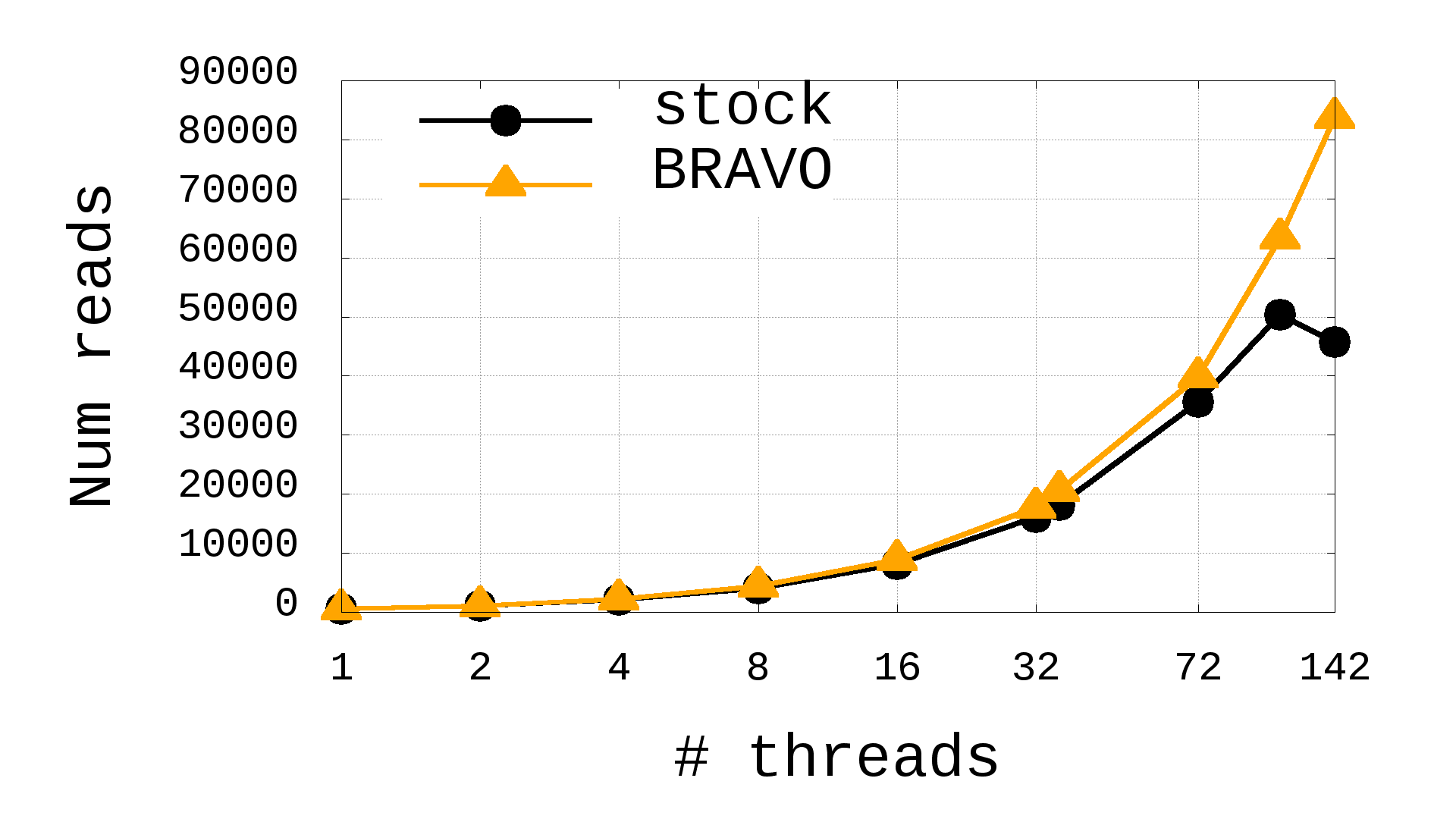}
\caption{readers ops}                                                   
\end{subfigure}%
~
\begin{subfigure}{0.25\textwidth}                                                                   
\centering
\includegraphics[width=1\textwidth]{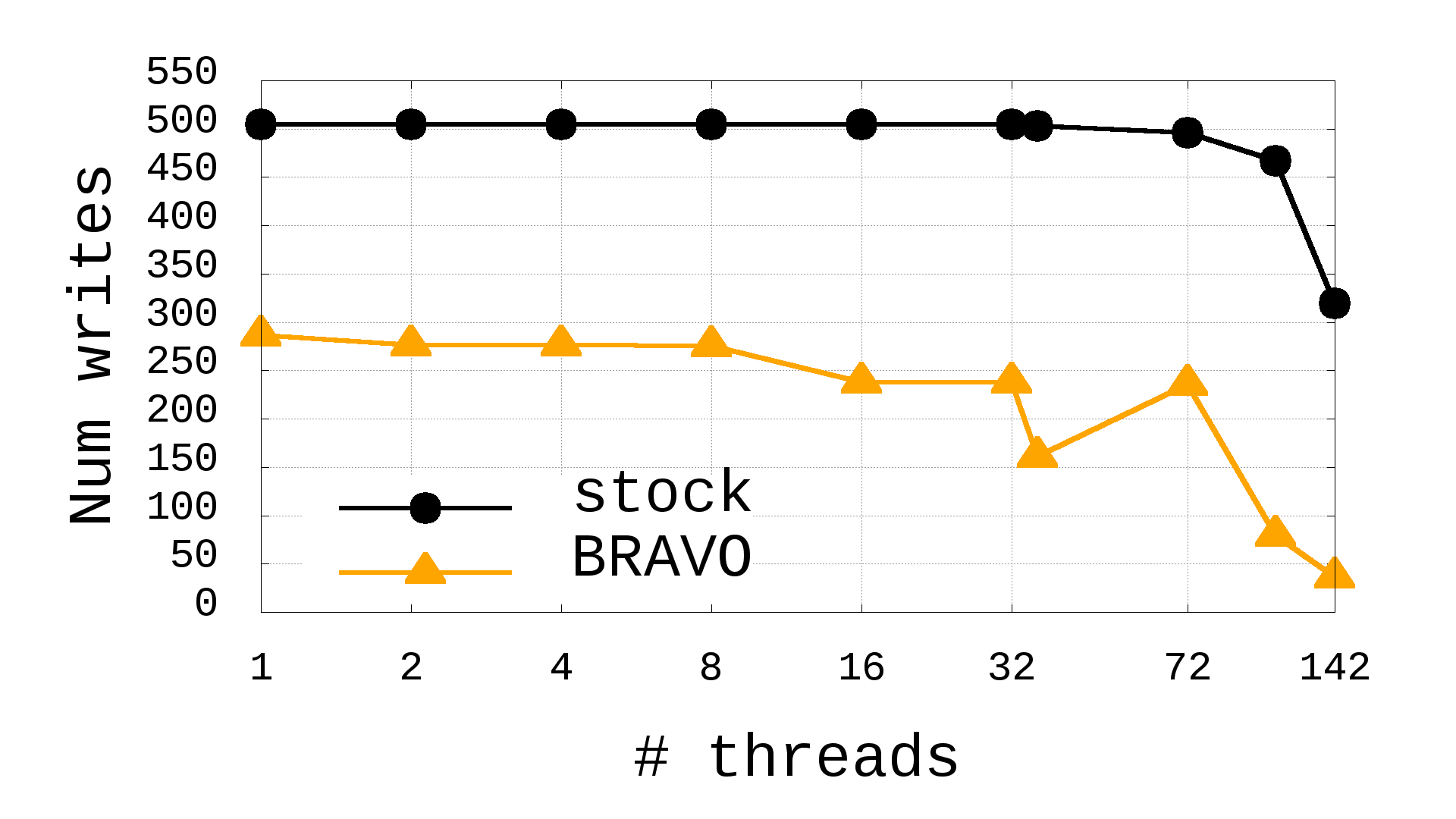}
\caption{writer ops}                                                   
\end{subfigure}%

\caption{\texttt{locktorture} results with 1 writer.}
\label{fig:locktorture-1-writer}
\end{figure}

Figure~\ref{fig:locktorture-1-writer} presents results for the experiment in which we vary the number of readers and set the number of writers to $1$.
We run the experiment for $30$ seconds, and report separately the total number of read and write acquisitions.
For low thread counts, the number of read operations on the stock and the BRAVO versions increases linearly in the number of threads.
Once the number of threads increases, the scalability of the stock is hampered.
The BRAVO version continues to scale perfectly across all thread counts.

When considering the number of write acquisitions, we note that the stock version has a better result.
This can be attributed to the nature of the benchmark, where a writer repeatedly acquires 
and releases the \texttt{rwsem} in the write mode.
In the BRAVO version, each such acquisition is likely to go through the revocation of the fast path,
where a writer waits for readers, which in this case have a relatively long critical section.
This increases the latency of the write acquisition and results in a smaller number of such acquisitions in the given period of time.
We validated this hypothesis by manually disabling the setting of \texttt{RBias} flag in BRAVO (that is, in that 
modified version, a fast acquisition path and revocation were never used), and observing that the modified BRAVO version
produced similar results to stock.

\begin{figure}

\begin{subfigure}{0.25\textwidth}                                                                   
\centering
\includegraphics[width=1\textwidth]{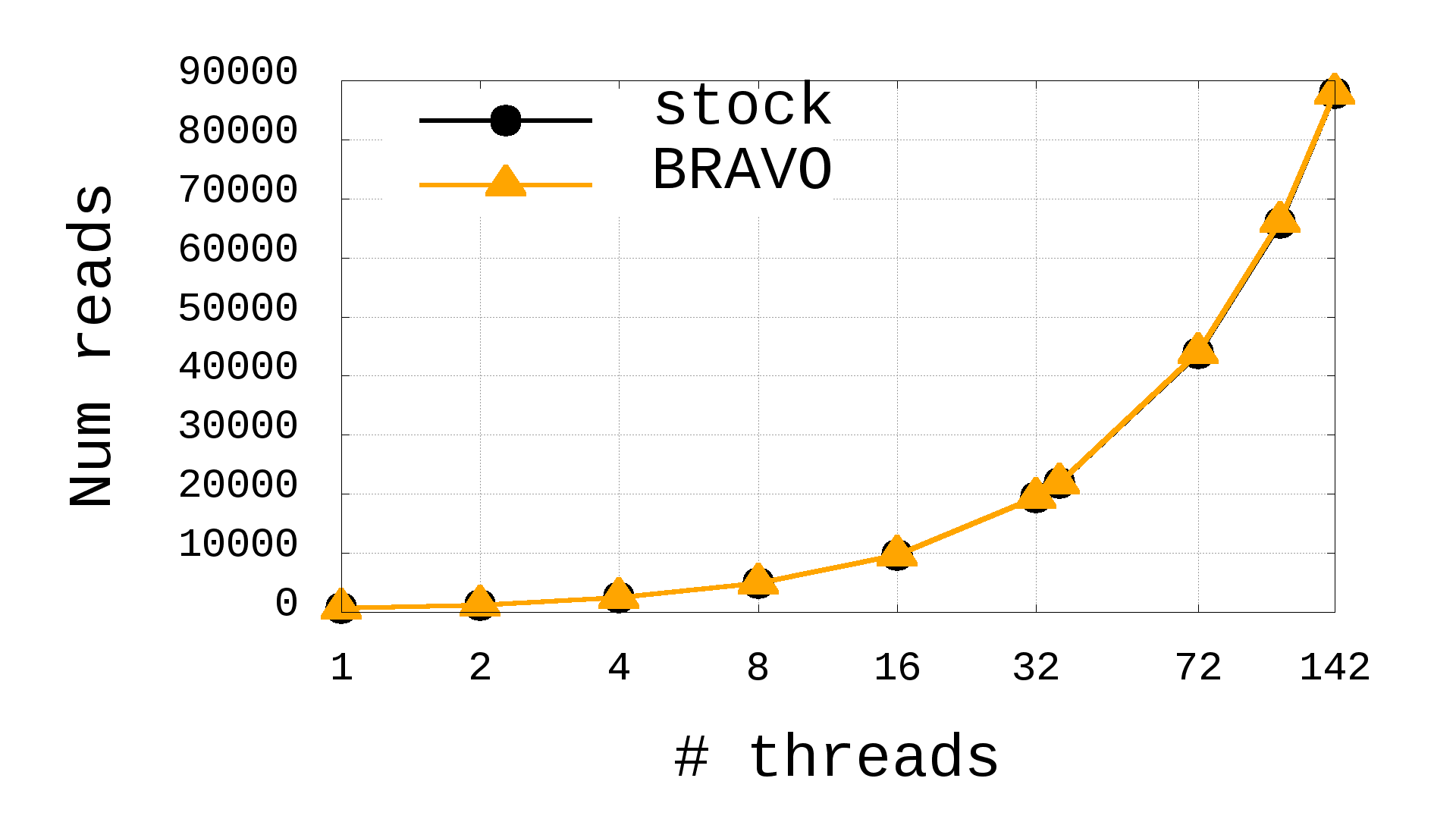}
\caption{original (50ms critical section)}                                                   
\end{subfigure}%
~
\begin{subfigure}{0.25\textwidth}                                                                   
\centering
\includegraphics[width=1\textwidth]{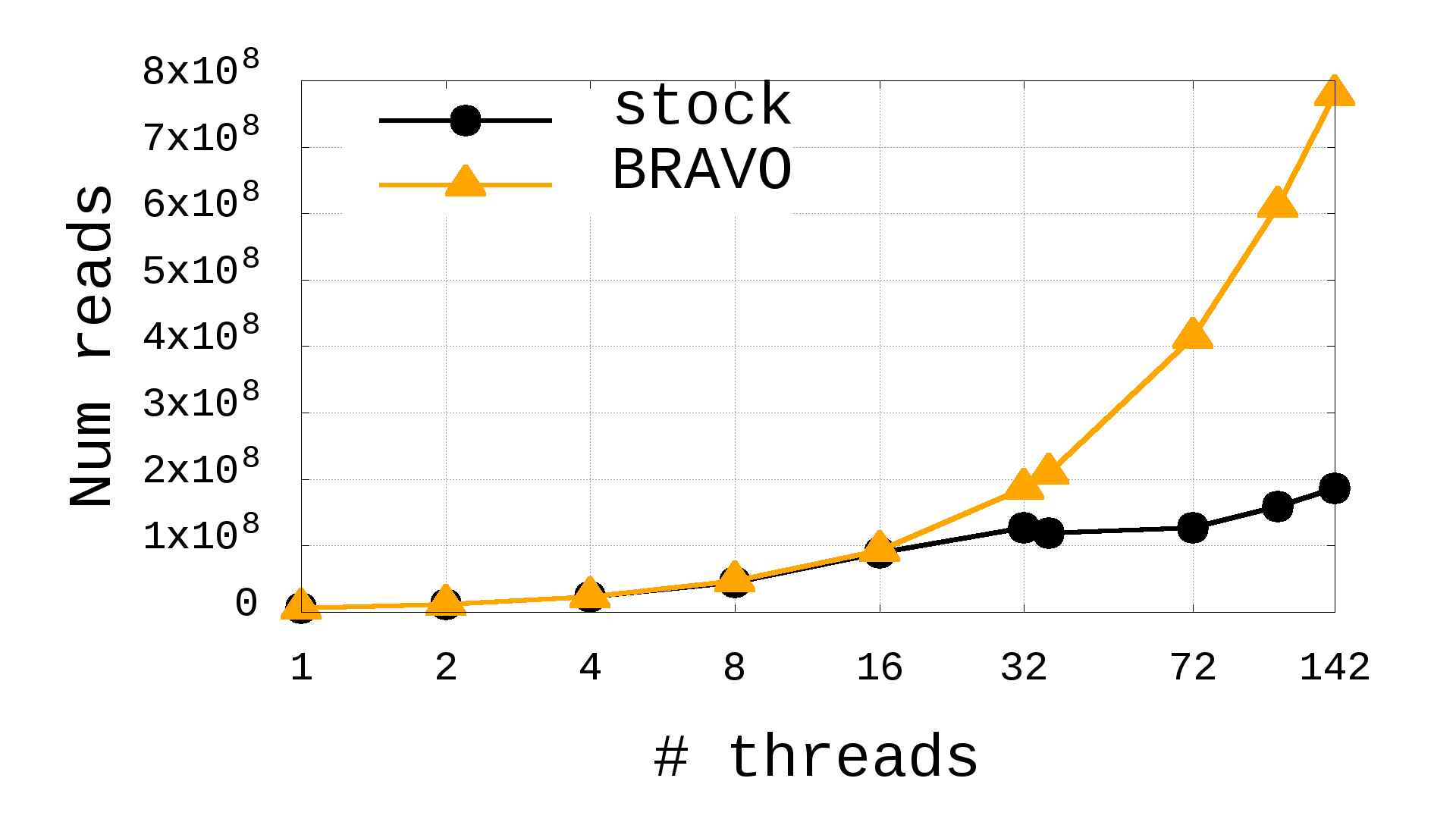}
\caption{modified (5us critical section)}                                                   
\end{subfigure}%

\caption{\texttt{locktorture} results with 0 writers.}
\label{fig:locktorture-0-writers}
\end{figure}

Figure~\ref{fig:locktorture-0-writers}~(a) presents results for the same experiment but with no writers.
Here both versions, stock and BRAVO, increase the number of reads linearly with the number of threads.
This is not surprising, given the relatively long read critical section (50ms), which masks any contention on the shared 
counter state during semaphore acquisition in the stock version.
To validate this claim, we modified the \texttt{locktorture} benchmark such that a reader would hold the lock for only 5us.
We note that we chose this number based on the typical length of a \texttt{rwsem} critical section as reported in 
\texttt{will-it-scale} benchmarks described in the next section.
We also made sure that readers in \texttt{locktorture} do not contend on any other shared variables, e.g., 
use a local random number generator seed instead of a shared one.
The results from this modified setup and a read-only workload are presented in Figure~\ref{fig:locktorture-0-writers}~(b).
As expected, they show that the stock version stops scaling once the contention on shared counter in \texttt{rwsem} grows.
At the same time, BRAVO avoids access to the shared counter and scales perfectly across all thread counts.
As a result, BRAVO refutes the common opinion that read-write locks should be used only when critical sections are long~\cite{CB08}.

\subsection{will-it-scale}
\texttt{will-it-scale} is an open-source collection of microbenchmarks\footnote{\url{https://github.com/antonblanchard/will-it-scale}} 
for stress-testing various kernel subsystems.
\texttt{Will-\allowbreak{}it-\allowbreak{}scale} runs in user-mode but is known to induce contention on kernel locks~\cite{DK19}. 
Each microbenchmark runs a given number of tasks (that can be either threads or processes), performing a series of specific system calls 
(such as opening and closing a file, mapping and unmapping memory pages, raising a signal, etc.).
We experiment with a subset of microbenchmarks that access the VMA structure and create contention on
\texttt{mmap\_sem}, an instance of \texttt{rwsem} that protects the access to VMA~\cite{Cor18}.
In particular, the relevant microbenchmarks are \texttt{page\_fault} and \texttt{mmap}.
The former continuously maps a large (128M) chunk of memory, writes one word into every page in the chunk 
(causing a page fault for every write), and unmaps the memory.
The latter simply maps and unmaps large chunks of memory.
(Each of those benchmarks
has several variants denoted as \texttt{page\_fault1}, \texttt{page\_fault2}, etc.)

Page faults require the acquisition of \texttt{mmap\_sem} for read, while memory mapping and unmapping operations acquire 
\texttt{mmap\_sem}s for write~\cite{CKZ12}.
Therefore, the access pattern for \texttt{mmap\_sem} is expected to be read-heavy in the \texttt{page\_fault} microbenchmark and 
more write-heavy in \texttt{mmap}.
We confirmed that through \texttt{lockstat} statistics.
We note that BRAVO is not expected to provide any benefit for \texttt{mmap}, yet we include it to evaluate 
any overhead BRAVO might introduce in write-heavy workloads.

\begin{figure}
\centering

\begin{subfigure}{0.25\textwidth}                                                                   
\centering
\includegraphics[width=1\textwidth]{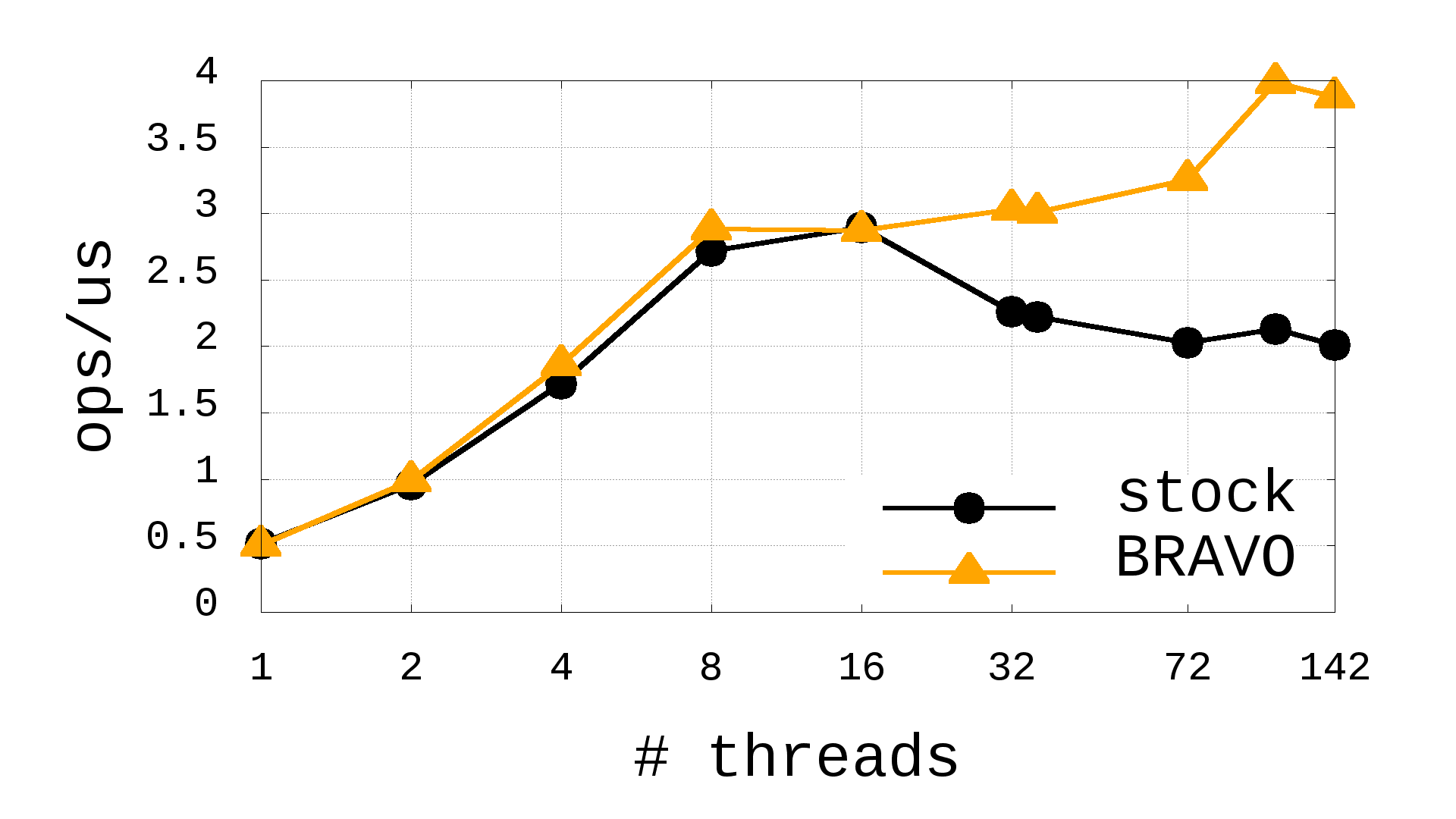}
\caption{page\_fault1\_threads}                                                   
\end{subfigure}%
~
\begin{subfigure}{0.25\textwidth}                                                                   
\centering
\includegraphics[width=1\textwidth]{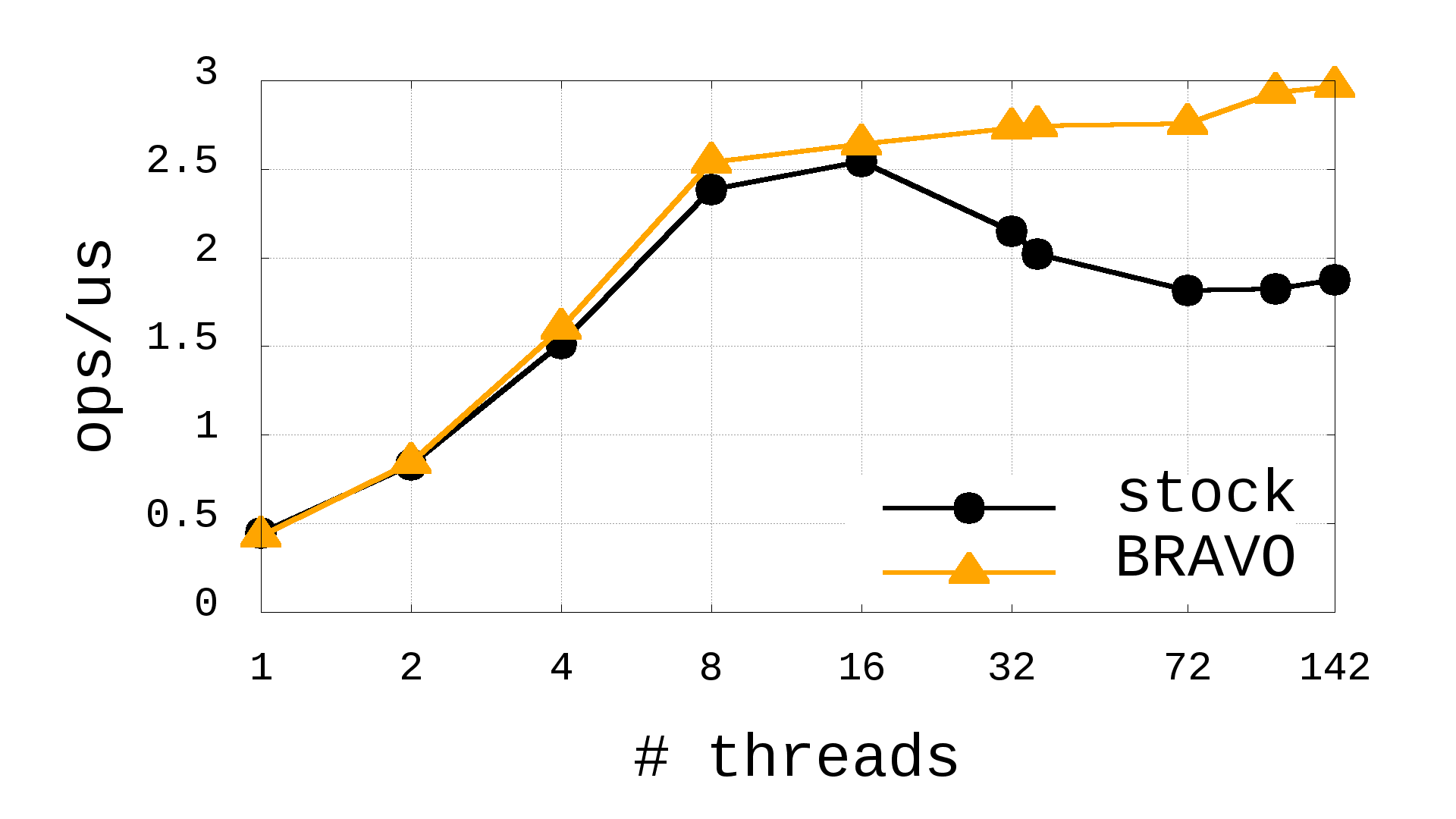}
\caption{page\_fault2\_threads}                                                   
\end{subfigure}%

\begin{subfigure}{0.25\textwidth}                                                                   
\centering
\includegraphics[width=1\textwidth]{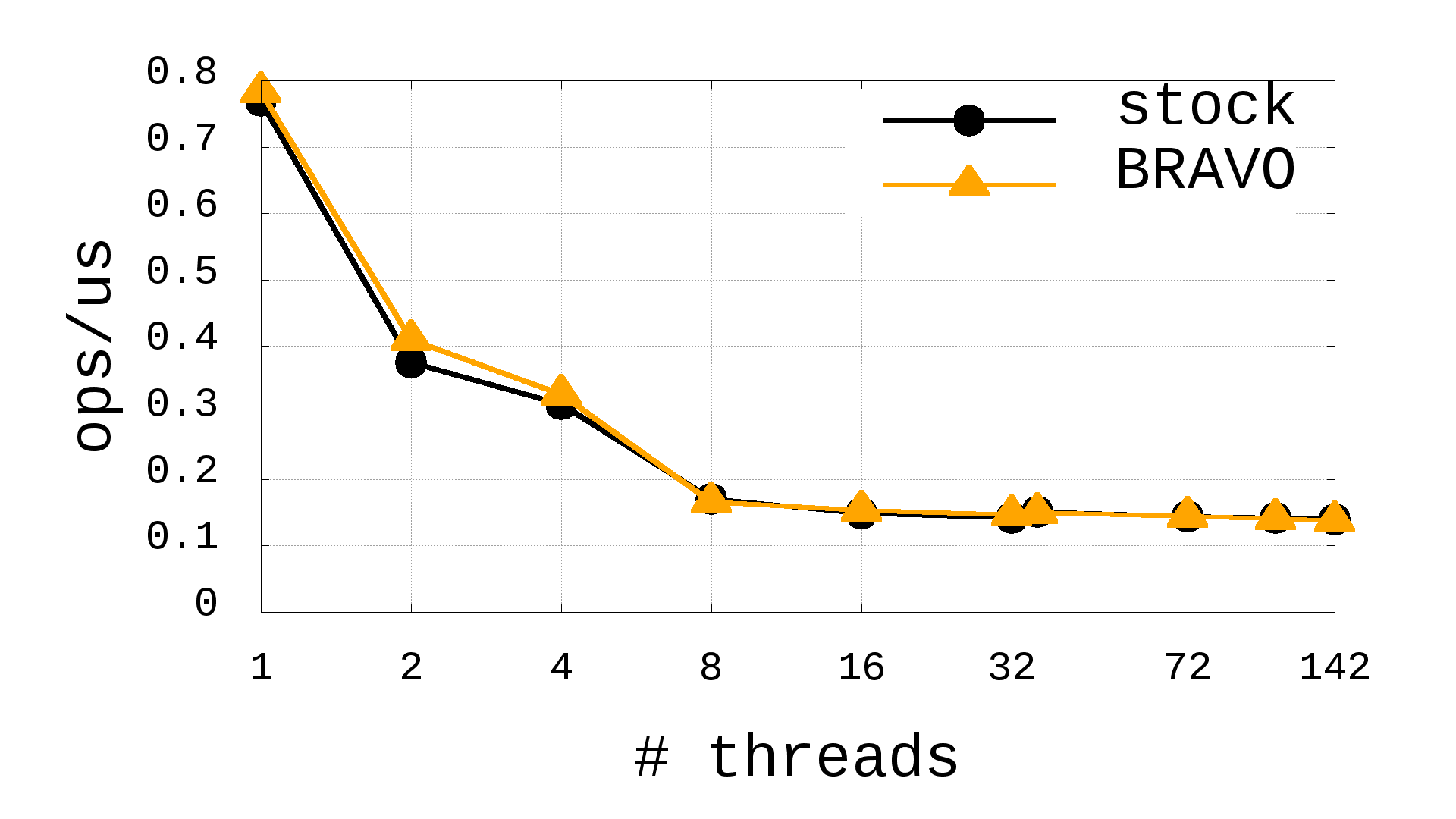}
\caption{mmap1\_threads}                                                   
\end{subfigure}%
~
\begin{subfigure}{0.25\textwidth}                                                                   
\centering
\includegraphics[width=1\textwidth]{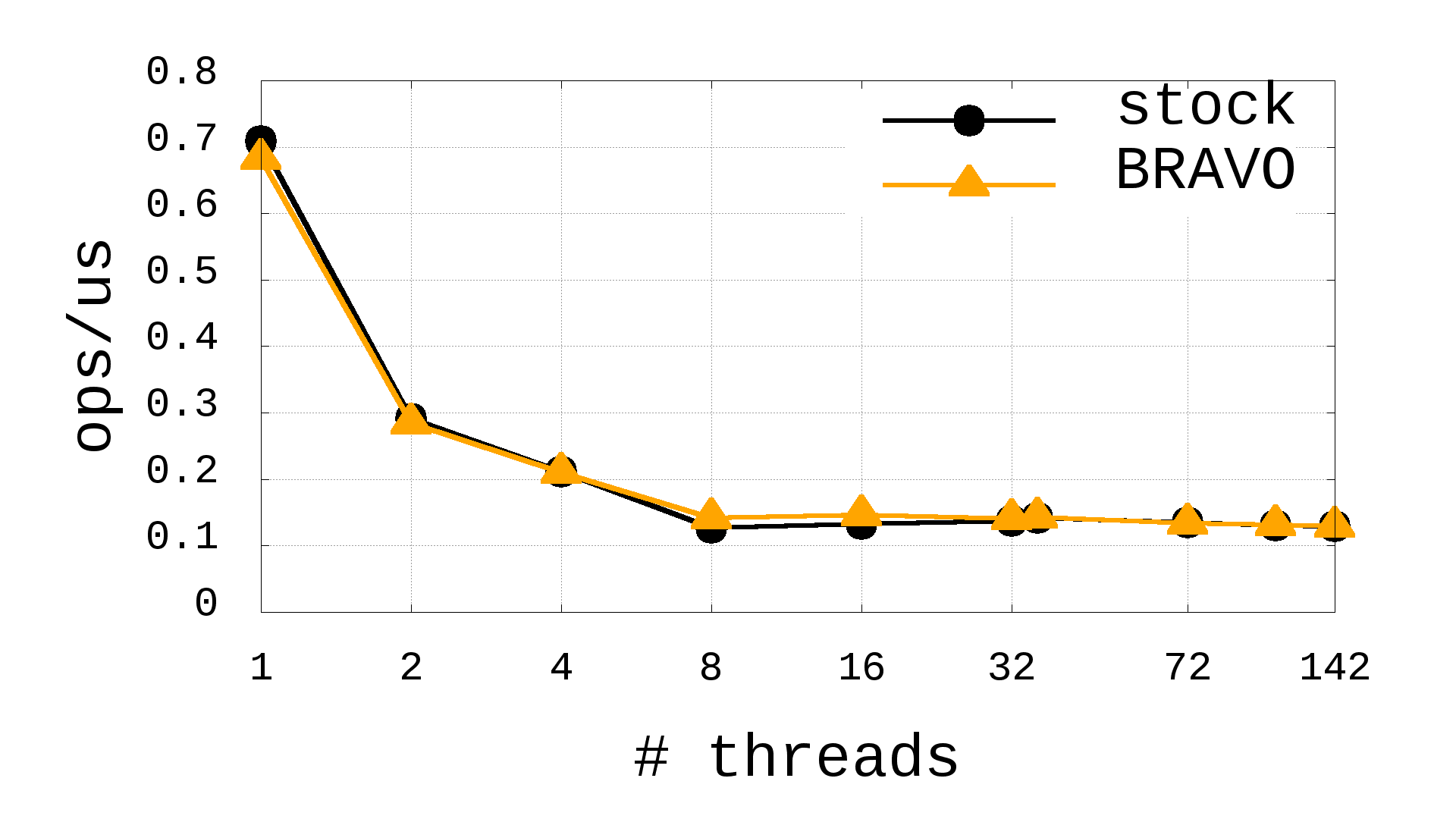}
\caption{mmap2\_threads}                                                   
\end{subfigure}%

\caption{\texttt{will-it-scale} results}
\label{fig:will-it-scale}
\end{figure}

Figure~\ref{fig:will-it-scale} presents the results of our experiments for \texttt{page\_fault} and \texttt{mmap}, respectively.
In \texttt{page\_fault}, the BRAVO version performs similarly to stock as long as the latter scales.
After $16$ threads, however, the throughput of the stock version decreases while the BRAVO version continues to scale, 
albeit at a slower rate.
At $142$ threads, BRAVO outperforms stock by up to $93\%$.
At the same time, \texttt{mmap} shows no significant difference in the performance of BRAVO vs. stock, 
suggesting that BRAVO does not introduce overhead in scenarios where it is not profitable.

\subsection{Metis}
Metis is an open-source MapReduce library~\cite{MMK10} used in the past to assess the scalability 
of Linux kernel locks~\cite{BWC10, CKZ12, usenixatc17-kashyap}.
Metis is known for a relatively intense access to VMA through the mix of page-fault and mmap operations~\cite{usenixatc17-kashyap}.
By collecting lock performance statistics with \texttt{lockstat}, however, we found that 
only few of Metis benchmarks have both a large number of \texttt{mmap\_sem}
acquisitions and a large portion of those acquisitions is for read.
We note that like in all other kernel benchmarks, \texttt{lockstat} was disabled when measuring performance numbers reported below.

Tables~\ref{table:metis-wc} and~\ref{table:metis-rwsem} present the performance results, respectively,
for \texttt{wc}, a map-reduce based word count, and \texttt{wrmem}, which allocates a large chunk of memory 
and fills it with random ``words'', which are fed into the map-reduce framework for inverted index calculation.
The BRAVO version can achieve speedups of over $30\%$.
We note that some of the data, particularly for \texttt{wc}, was noisy; we print values with variance larger 
than $5\%$ from the mean in italics.
(All values have variance of $19\%$ or less).
We also note that BRAVO did not create significant overhead for any other Metis benchmark, although some 
benchmarks produced noisy results similarly to \texttt{wc}.

\begin{table}[t]
\begin{center}
\begin{tabular}{ccccc}
\#threads & stock & BRAVO & speedup\\
\hline
  1  & \emph{18.059} & 17.957 & 0.6\%\\
  2  & 12.189 & 12.090 & 0.8\%\\
  4  & 10.045 & \emph{9.652} & 3.9\%\\
  8  & \emph{8.880} & \emph{7.976} & 10.2\%\\
 16  & \emph{13.321} & 11.325 & 15.0\%\\
 32  & 23.654 & \emph{19.281} & 18.5\%\\
 72  & \emph{119.018} & 98.054 & 17.6\%\\
108  & 132.147 & 111.139 & 15.9\%\\
142  & 143.217 & 125.160 & 12.6\%\\
\hline
\end{tabular}
\end{center}
\caption{\texttt{wc} runtime (sec)}
\label{table:metis-wc}
\end{table}

\begin{table}[t]
\begin{center}
\begin{tabular}{ccccc}
\#threads & stock & BRAVO & speedup\\
\hline
  1  & 279.553 & 279.423 & 0.0\%\\
  2  & 137.271 & 136.706 & 0.4\%\\
  4  & 68.989 & 69.013 & 0.0\%\\
  8  & 36.161 & 36.210 & -0.1\%\\
 16  & 22.647 & 21.985 & 2.9\%\\
 32  & 18.332 & 14.707 & 19.8\%\\
 72  & \emph{48.397} & 31.868 & 34.2\%\\
108  & 58.529 & \emph{36.910} & 36.9\%\\
142  & 58.994 & 42.544 & 27.9\%\\
\hline
\end{tabular}
\end{center}
\caption{\texttt{wrmem} runtime (sec)}
\label{table:metis-rwsem}
\end{table}

\section{Conclusion and Future Work}
\label{sec:conclusions} 

BRAVO easily composes with existing locks, preserving desirable properties of those
underlying locks, and yielding a composite lock with improved read-read scalability.
We specifically target read-dominated workloads with multiple concurrent threads
that acquire and release read permission at a high rate.  
The approach is simple, effective, and yields improved
performance for read-dominated workloads compared to commonly used compact locks.
The key trade-off inherent in the design is the benefit accrued by reads against the 
potential slow-down imposed by revocation.  Even in mixed or write-heavy workloads, we limit any 
slow-down stemming from revocation costs and bound harm, making the decision to use 
BRAVO simple.  BRAVO incurs a very small footprint increase per lock instance, and also 
adds a shared table of fixed size that can be used by all threads and locks.  
BRAVO's key benefit arises from reducing coherence cost that would normally be
incurred by locks having a central reader indicator.  Write performance is left unchanged relative
to the underlying lock.  BRAVO provides read-read performance at, and often above, that of the
best modern reader-writer locks that use distributed read indicators, but without the footprint
or complexity of such locks.  By reducing coherence traffic, BRAVO is implicitly NUMA-friendly.

\AtFoot{An extended version of this paper is available at \url{https://arxiv.org/abs/1810.01573}} 

\Boldly{Future directions}
We identify a number of future directions for our investigation into BRAVO-based designs:
\begin{itemize}[leftmargin=4mm] 
\item Dynamic sizing of the visible readers table based on collisions. 
Large tables will have reduced collision rates but larger scan revocation overheads.
\item The reader fast-path currently probes just a single location and reverts to the slow-path
after a collision.  We plan on using a secondary hash to probe an alternative location.
In that vein, we note that while we currently use a hash function to map a thread's identity and the lock address to an
index in the table, there is no particular requirement that the function that associates a read request
with an index be deterministic.  We plan on exploring other functions, using time or random numbers
to form indices. While this will be less beneficial in terms of cache locality for the reader, it might be helpful
in case of temporal contention over specific slots.
\item Accelerate the revocation scan operation via SIMD instructions such as AVX.  The visible reader
table is usually sparsely populated, making it amenable to such optimizations.   Non-temporal
non-polluting loads may also be helpful for the scan operation.  
\item As noted, our current policy to enable bias is conservative, and leaves
untapped performance.  We intend to explore more sophisticated adaptive policies based on recent behavior
and to use a more faithful cost model. 
\item An interesting variation is to implement BRAVO on top of an underlying mutex
instead of a reader-writer lock.  Slow-path readers must acquire the mutex, and the sole source
of read-read concurrency is via the fast path.  We note that some applications might expect
the reader-write lock implementation to be fully work conserving and \emph{maximally admissive} -- always allowing full
read concurrency where available.  For example an active reader thread $T1$, understanding by virtue of application invariants  
that no writers are present, might signal another thread $T2$ and expect that $T2$ can enter a reader critical section while 
$T1$ remains within the critical section.  This progress assumption would not necessarily hold if readers are forced
through the slow path and read-read parallelism is denied.  
\Invisible{If no writers are waiting or arrive, and a reader arrives while another reader is active, the
first reader will be allowed admission -- if a thread \emph{can} enter, then it \emph{will} enter.} 
\item In our current implementation arriving readers are blocked while a revocation scan
is in progress.  This could be avoided by adding a mutex to each BRAVO-enhanced lock.  Arriving writers
immediately acquire this mutex, which resolves all write-write conflicts.  They then perform revocation,
if necessary; acquire the underlying reader-vs-write lock with write permission; execute the writer 
critical section; and finally release both the mutex and the underlying reader-writer lock.
The underlying reader-writer lock resolves read-vs-write conflicts.  
The code used by readers remains unchanged.  
This optimization allows readers to make progress during revocation by diverting through the
reader slow-path, mitigating the cost of revocation.  
This also reduces variance for the latency of read operations. 
We note that this general technique can be readily applied to other existing reader-writer locks
that employ distributed reader indicators, such as Linux's \emph{br-lock} family \cite{usenixatc14-liu}.  
\item Modify the hash function that selects indices so that a given lock maps to subsets 
of the visible readers table.  In turn, we can then accelerate the revocation scan by restricting the
scan to those subsets.   
\item Partition the visible readers table into contiguous sectors specified by the NUMA node of the reader.  
This optimization will reduce the cost of false sharing within the table.   
\item 
Our preferred embodiment is \textbf{BRAVO-2D} where we partition the visible reader table into disjoint ``rows''.    
A row is a contiguous set of slots, and is aligned on a cache sector boundary.
We configure the length of a row as a multiple of the cache sector size.  
Readers index into the table by using the caller's CPUID to select a row.   
and then hash the lock address to form an offset into that row.   
We could think of this offset as selecting a column.   
This approach virtually eliminates inter-thread interference and false sharing (near collisions).
Threads enjoy spatial and temporal locality within their row.  
Inter-thread collisions would typically arise only because of migration and preemption, which we expect to be rare. 
The thread-to-CPUID relationship is expected to be fairly stable over the short term.

We may still encounter inter-lock collisions within a row, but fall back as usual to the underlying lock.
Note that inter-lock collisions will be intra-thread, within a row.  
BRAVO-2D admits a higher inter-lock collision rate than does our baseline approach, where we just hash the 
thread and lock to form the index.   Specifically, for a given thread and lock, the set of possible
slots is restricted to just one row under BRAVO-2D, instead of the how table as was the case in the 
baseline design.  
But in practice we find that improved intra-row locality in the BRAVO-2D form overcomes that penalty.   
And since most threads hold very few reader-writer locks at a given time, the odds of inter-lock collisions
are still very low, even if slightly higher compared to our baseline.   
At revocation time we need only scan the column associated with the lock, instead of the full table,
reducing the number of slots accessed compared to the baseline design and accelerating revocation. 
Furthermore the automatic hardware stride-based prefetcher can still track the revocation access pattern 
and provide benefit.  
\end{itemize}

\begin{acks}                            

We thank Shady Issa for useful discussions about revocation and the cost model.
We also thank the anonymous reviewers and our shepherd Yu Hua for providing insightful comments.


\end{acks}

\bibliography{BRAVO.bib}



\end{document}